\documentclass[manuscript]{acmart}

\AtBeginDocument{%
  \providecommand\BibTeX{{%
    Bib\TeX}}}

\usepackage{amsmath}
\usepackage{amsthm}
\usepackage{algorithmic}
\usepackage{graphicx}
\usepackage{textcomp}
\usepackage{xcolor}
\usepackage{multirow}
\usepackage{booktabs}
\usepackage{pifont}

\usepackage[linesnumbered,ruled,vlined]{algorithm2e}
\usepackage[utf8]{inputenc}
\usepackage{mathrsfs}
\usepackage{listings}
\usepackage{enumitem}
\usepackage{url}

\usepackage{array}
\usepackage{mathtools}
\usepackage{fancyvrb}
\usepackage{makecell}
\usepackage{colortbl}
\usepackage{siunitx}
\usepackage{tikz}
\newcommand{\circled}[1]{%
  \tikz[baseline=(char.base)]{
    \node[shape=circle,draw,inner sep=1pt] (char) {#1};
  }%
}

\definecolor{myblue}{RGB}{0,100,200}

\renewcommand{\arraystretch}{1.4}
\setlist[itemize]{noitemsep, topsep=0pt}

\definecolor{myred}{rgb}{0.8, 0.1, 0.1}

\newtheorem{definition}{Definition}

\usepackage{verbatim}

\newcommand{\yfp}[1]{{\color{blue} #1}}
\newcommand{\oomit}[1]{}

\newcommand{\toolname}{SESpec}

\definecolor{GPTGreen}{rgb}{0.459,0.675,0.616}
\definecolor{Green}{rgb}{0,0.5,0}
\definecolor{Blue}{rgb}{0.035, 0.682, 0.855}
\definecolor{gray}{rgb}{0.5, 0.5, 0.5}

\def\BibTeX{{\rm B\kern-.05em{\sc i\kern-.025em b}\kern-.08em T\kern-.1667em\lower.7ex\hbox{E}\kern-.125emX}}

\begin{document}

\title{Integrating Symbolic Execution with LLMs for Automated Generation of Program Specifications}


\author{Fanpeng Yang}
\affiliation{
  \institution{Institute of Software, Chinese Academy of Sciences, UCAS}
  \city{Beijing}
  \country{China}
}
\email{yangfp@ios.ac.cn}

\author{Xu Ma}
\affiliation{
  \institution{Institute of Software, Chinese Academy of Sciences, UCAS}
  \city{Beijing}
  \country{China}
}
\email{maxu@ios.ac.cn}

\author{Shuling Wang}
\authornote{Corresponding author}
\affiliation{
  \institution{Institute of Software, Chinese Academy of Sciences, UCAS}
  \city{Beijing}
  \country{China}
}
\email{wangsl@ios.ac.cn}

\author{Xiong Xu}
\affiliation{
  \institution{Institute of Software, Chinese Academy of Sciences}
  \city{Beijing}
  \country{China}
}
\email{xux@ios.ac.cn}

\author{Qinxiang Cao}
\affiliation{
  \institution{Sch. of Computer Science \& Sch. of Artificial Intelligence, Shanghai Jiao Tong University}
  \city{Shanghai}
  \country{China}
}
\email{caoqinxiang@sjtu.edu.cn}

\author{Naijun Zhan}
\affiliation{
  \institution{School of Computer Science, Peking University}
  \city{Beijing}
  \country{China}
}
\email{znj@ios.ac.cn}

\author{Xiaofeng Li}
\affiliation{
  \institution{Beijing Institute of Control Engineering}
  \city{Beijing}
  \country{China}
}
\email{1351127502@126.com}

\author{Bin Gu}
\affiliation{
  \institution{Beijing Institute of Control Engineering}
  \city{Beijing}
  \country{China}
}
\email{gubin@ios.ac.cn}

\renewcommand{\shortauthors}{Yang et al.}


\begin{abstract}
Automatically generating formal specifications including loop invariants, preconditions, and postconditions for legacy code is critical for program understanding, reuse and verification. However, the inherent complexity of control and data structures in programs makes this task particularly challenging. This paper presents a novel framework that integrates symbolic execution with large language models (LLMs) to automatically synthesize formally verified program specifications. Our method first employs symbolic execution to derive precise strongest postconditions for loop‑free code segments. These symbolic execution results, along with automatically generated invariant templates, then guide the LLM to propose and iteratively refine loop invariants until a correct specification is obtained. The template‑guided generation process robustly combines symbolic inference with LLM reasoning, significantly reducing hallucinations and syntactic errors by structurally constraining the LLM's output space. Furthermore, our approach can produce strong specifications without relying on externally provided verification goals, enabled by the rich semantic context supplied by symbolic execution, overcoming a key limitation of prior goal‑dependent tools. Extensive evaluation shows that our tool \toolname\ outperforms the existing state-of-the-art tools across numerical and data-structure benchmarks, demonstrating both high precision and broad applicability.
\end{abstract}

\ccsdesc[500]{Theory of computation~Automated reasoning}
\ccsdesc[500]{Software and its engineering~Software functional properties}
\ccsdesc[500]{Computing methodologies~Artificial intelligence}
\keywords{Program specification, Loop invariants, Symbolic execution, Large language models, Formal verification}


\maketitle

\section{Introduction}

 Across diverse software domains, ranging from general-purpose systems to safety-critical fields such as aerospace, a vast amount of legacy code exists but lacks formal specifications defining its functionality. Automatically generating program specifications is therefore essential for program comprehension, reuse, and formal verification against the design requirements. Common forms of formal specifications include loop invariants, preconditions, and postconditions. However, generating such specifications remains challenging due to complex control structures (e.g., nested branches, loops,  and function calls) and intricate data structures (e.g., structs, pointers,  and inductive datatypes such as linked lists).

In this work, we focus on the problem of functional correctness specification. Our goal is to generate specifications that fully characterize program behavior across all execution paths and relevant variables, rather than producing the specifications  sufficient only to satisfy a particular verification goal. We therefore view specification generation as a form of functional summarization. While verification goals play an important role and can guide the generation process, our approach is not limited to producing specifications tailored only to a given goal. Instead, we seek to infer specifications that are as strong and informative as possible. As a principled first step, we ground the specification generation process by leveraging precise, path-sensitive symbolic execution.

Existing research  on program specification generation has predominantly centered around loop invariants. Traditional techniques reduce the problem of synthesizing loop invariants to constraint solving~\cite{DBLP:conf/cav/ColonSS03,DBLP:journals/pacmpl/LiuFYSL22}, Craig interpolation~\cite{FiB_Squeezing_loop,DBLP:conf/cade/GanDXZKC16,Nonlinear_Craig_Interpolant_Generation}, recurrence analysis~\cite{beyer2024decomposing,wonisch2012predicate}, shape analysis~\cite{calcagno2009compositional}, \emph{etc},  however, these approaches suffer from restrictions in the types of programs and invariants they can handle; moreover, their complexity often hinders scalable, automated inference, requiring manual guidance. Dynamic methods, such as Daikon~\cite{daikon}, synthesize invariants by monitoring test executions against predefined templates, however, their effectiveness heavily relies on test coverage and quality, and they provide no formal guarantees of correctness. Recent efforts combine machine learning with SMT solvers to improve efficiency~\cite{Code2Inv_a_Deep_Learning_Framework_for_Program_Verification,Loop_Invariant_Inference_through_SMT_Solving_Enhanced_Reinforcement_Learning,DBLP:journals/pacmse/CaoWXYWCM25}, but these methods remain largely confined to numeric programs. More recently, Large Language Models (LLMs) have been applied to the specification generation of broader programs.  While LLMs exhibit a strong capability to understand program semantics, they often struggle to produce syntactically valid and semantically precise formal specifications: hallucinations may lead to ill-formed assertions, and even syntactically correct specifications tend to be incorrect or overly trivial, failing to faithfully capture program behavior.

Several recent efforts have sought to address these challenges ~\cite{autospec,Enhancing_Automated_Loop_Invariant_Generation_for_Complex_Programs_with_Large_Language_Models,SpecGen}. Among them, AutoSpec, the state-of-the-art LLM-based approach, integrates static analysis, LLMs, and formal verification to decompose programs and generate specifications bottom-up. However, its performance degrades for programs involving data structures such as structs and linked lists. More fundamentally, while static analysis enables a divide-and-conquer decomposition of programs, the specification generation of AutoSpec still largely relies on LLMs’ semantic understanding. As a result, static analysis alone is insufficient to mitigate hallucinations at both the syntactic and semantic levels. Finally, like most existing approach, AutoSpec is goal-directed, relying on predefined verification goals to synthesize specifications sufficient to prove a given property. Such specifications are often narrowly tailored to a single objective and may over-approximate the behaviors outside that goal, thus failing to describe a program’s overall functional correctness, particularly when explicit verification goals are absent.

To overcome these limitations, we propose a novel method for generating program specifications even in the absence of explicit verification goals, with the primary objective of capturing a program’s functional correctness rather than merely proving a specific property. Our approach integrates symbolic execution,  LLMs and formal verification to generate formally verified specifications. This unified methodology handles diverse program constructs in a principled manner.
For loop-free code, symbolic execution precisely tracks path conditions and state transitions, directly constructing strongest specifications that characterize exact functional behavior. For loops, candidate invariants are generated by LLMs within localized contexts guided by precise symbolic execution results and equality-constraint based invariant templates. These invariants are then iteratively refined using verification feedback and adaptive LLM strategies until they are formally validated, and are sufficiently strong to support verification goals when such goals are
provided. For complex constructs, they  are handled in a bottom-up fashion. Function calls and nested loops are managed using stacks that enable compositional specification generation from inner to outer scopes. Non-recursive data structures, including pointers, arrays, and structs, are modeled using a flattened memory representation that explicitly tracks separation and aliasing, while recursive data structures are specified using inductive predicates. Together, these techniques are designed to collectively infer maximally precise functional correctness specifications.

In the actual implementation, we leverage QCP's symbolic execution engine~\cite{wu2025qcppracticalseparationlogicbased,vst-a} for its ability to precisely track symbolic execution results at arbitrary execution points, combined with Frama-C~\cite{cuoq2012frama} as our formal verifier due to its automated verification capability and also ACSL-style assertions that are  amenable to LLM processing. Our framework tightly integrates QCP, Frama-C and LLMs, where each component's strengths compensate for the others' limitations.

We performed a comprehensive and rigorous experimental evaluation to validate our approach.
This includes an ablation study confirming the contribution of each technical component, along with extensive benchmark assessments. The results demonstrate the effectiveness and broad applicability of our approach across diverse tasks in invariant and program specification generation. On numerical benchmarks, our method performs competitively with specialized tools designed for numerical programs.
Compared to recent LLM-based approaches such as AutoSpec and ACInv, our approach demonstrates superior performance on numerical benchmarks. 
Furthermore, we introduced three new benchmarks targeting distinct programs: more complex numerical programs, multi-layered structs and branches derived from real aerospace applications, and recursive data structures with singly linked lists. Our method significantly outperforms both AutoSpec and ACInv across  these benchmarks, demonstrating  the generalizability and robustness of our method across a wide spectrum of programs.
Additionally, we evaluated our method on loop invariant and program specification generation under masked verification goals using the SyGuS and Frama-C benchmarks. The results indicate that our approach continues to produce high-quality specifications with a small drop in performance, while substantially outperforming the state-of-the-art tool AutoSpec under identical masking conditions. These findings highlight the robustness  of our approach even in the absence of explicit verification goal guidance.


The rest of the paper is organized as follows: Sect.~\ref{sec:motivation} introduces a motivating example and provides an overview of our approach. Sect.~\ref{sec:summarygeneration} and Sect.~\ref{sec:loopinvariant} detail our core technical contributions on the generation of function specifications and loop invariants. Sect.~\ref{sec:experiments} evaluates our approach empirically. Sect.~\ref{sec:threats} discusses threats to validity. Finally, Sect.~\ref{sec:relatedwork} presents the related work, and Sect.~\ref{sec:conclusion} concludes the paper with directions for future work. The details on tool implementation, LLM configurations, and experimental results are available at \url{https://github.com/anon-hiktyq/TOSEM2026-SESpec}
.

\section{A Motivating Example And Our Approach}

\label{sec:motivation}

\newcommand{\annotate}[1]{{\color{blue} \mathsf{#1}}}

\begin{figure*}
    \centering\includegraphics[width=1.0\textwidth]{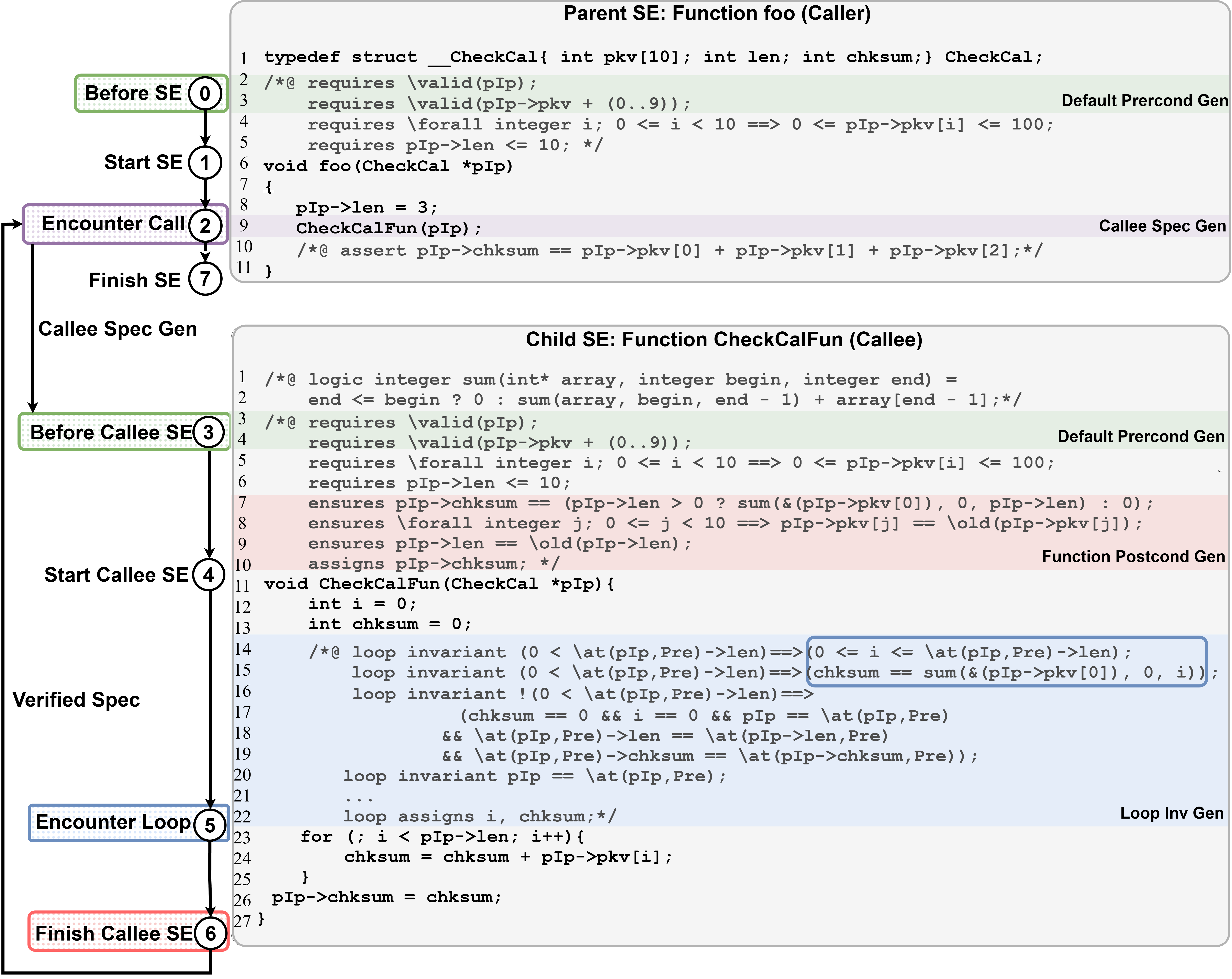}
    \caption{Motivating Example}
    \label{fig:example}
\end{figure*}

Automating the generation of formal specifications for real-world C code remains challenging due to the prevalence of complex language features such as loops, pointer manipulations and data structures. Consider the code fragment shown in Fig. \ref{fig:example}, which originates from the Sun Search system, a legacy embedded software component in the aerospace domain. The code starts to define a struct \texttt{CheckCal} with three fields: an integer array \texttt{pkv[10]},  an integer field \texttt{len} representing the number of valid elements in the array, and an integer \texttt{chksum} storing  the sum of the array's elements. Then, function \texttt{foo} invokes \texttt{CheckCalFun} on a \texttt{CheckCal} pointer \texttt{pIp}, with a given verification goal annotated by \texttt{/*@assert $\cdots$ */}.  The callee function \texttt{CheckCalFun} takes as input a \texttt{CheckCal} pointer $\texttt{pIp}$, computes the sum of  the $\texttt{len}$ elements in the array $\texttt{pkv}$ via a loop and then stores the result in the field \texttt{chksum} of \texttt{pIp}. 
This example exhibits diverse features: complex data structures (structs, arrays and pointers), a loop with inductive operations over arrays, and  pointer-induced aliasing and side effects.
In consequence, it faces the following challenges:
\begin{itemize}
  \item \textbf{Loop reasoning:}  
  The summation loop \texttt{for(; i<pIp->len; i++)} requires  determining whether the initial condition allows to enter the loop, and if the loop executes, establishing the invariants sufficient enough to specify the precise behavior of the loop, i.e. to relate the partial sum to the processed prefix of the array.
  
  \item \textbf{Structs and Pointers:}  The procedures involve operations such as accessing arrays, reading and writing to struct fields via pointers, making it essential to account for the resulting aliasing and side effects in their specifications. For instance, the specification for function \texttt{CheckCalFun} must guarantee memory safety for dereferencing \texttt{pIp->pkv[i]},  precisely describe  the write effect on \texttt{pIp->chksum}, and specify the frame property for other fields of \texttt{pIp} that remain unchanged. 
  
  \item \textbf{Inductive predicates:}  
  Properties of arrays in loop such as aggregate sums are  most naturally expressed with inductive predicates. Automatically writing correct forms of such predicates and furthermore verifying them remains a central challenge for programs involving recursive operations or data structures.
\end{itemize}

As shown in Fig.~\ref{fig:example}, the code is annotated with the full ACSL specifications enclosed in $/*@ ... */$, including preconditions (beginning with \texttt{requires}), postconditions (beginning with \texttt{ensures}), loop invariants (beginning with \texttt{loop\ invariant}),  and verification goals (beginning with \texttt{assert}). 
Among these, the loop invariants,   the default preconditions and the postconditions  are generated automatically by our tool \toolname, while the \texttt{assert} clause  represents verification goals provided a priori, and certain (non-default, line 4-5) preconditions reflect additional requirements from the designers.

Before presenting our method, we attempted to generate the specification for this example using GPT-4o, GPT-5 and AutoSpec respectively,  given both the preconditions and the verification goal. The loop invariants they produced, shown below, were all found to be incorrect:


\[
\small
\begin{aligned}
  & \texttt{loop invariant 0 <= i <= pIp->len;} \\
  & \texttt{loop invariant chksum == \textbackslash sum(0, i-1, \textbackslash lambda integer k; pIp->pkv[k]); \quad (\textbf{GPT-4o})} \\ 
  \\ 
  & \texttt{loop invariant 0 <= i <= pIp->len;} \\
  & \texttt{loop invariant chksum == \textbackslash sum(integer k = 0; k < i; pIp->pkv[k]);} \\
  & \texttt{loop assigns i, chksum;} \\
  & \texttt{loop variant pIp->len - i; \quad (\textbf{GPT-5})} \\ 
  \\ 
  & \texttt{loop invariant 0 <= i <= pIp->len;} \\
  & \texttt{loop invariant chksum == \textbackslash sum(0, i-1, pIp->pkv);} \\
  & \texttt{loop invariant \textbackslash forall integer k; 0 <= k < i ==> chksum >= pIp->pkv[k];} \\
  & \texttt{loop assigns i, chksum; \quad (\textbf{AutoSpec})}
\end{aligned}
\]

The failures highlight the common limitations of pure LLM-based approaches. First, LLMs often neglect whether a loop is ever entered, so their generated invariants presume loop entry without establishing necessary preconditions (e.g., \texttt{pIp->len} might be negative). Even when we explicitly prompted GPT-4o to "consider the case when the loop is not entered", it still failed to generate the correct conditional invariant structure. Second, when reasoning about arrays, LLMs introduce uninterpreted functions (e.g., a custom \texttt{sum}) that are undefined in the target verification framework (such as Frama-C), causing verification failures. 

\label{sec:architecture}

\subsection{Our Approach} 



\begin{figure*}[h]
    \centering
    \includegraphics[width=0.8\textwidth]{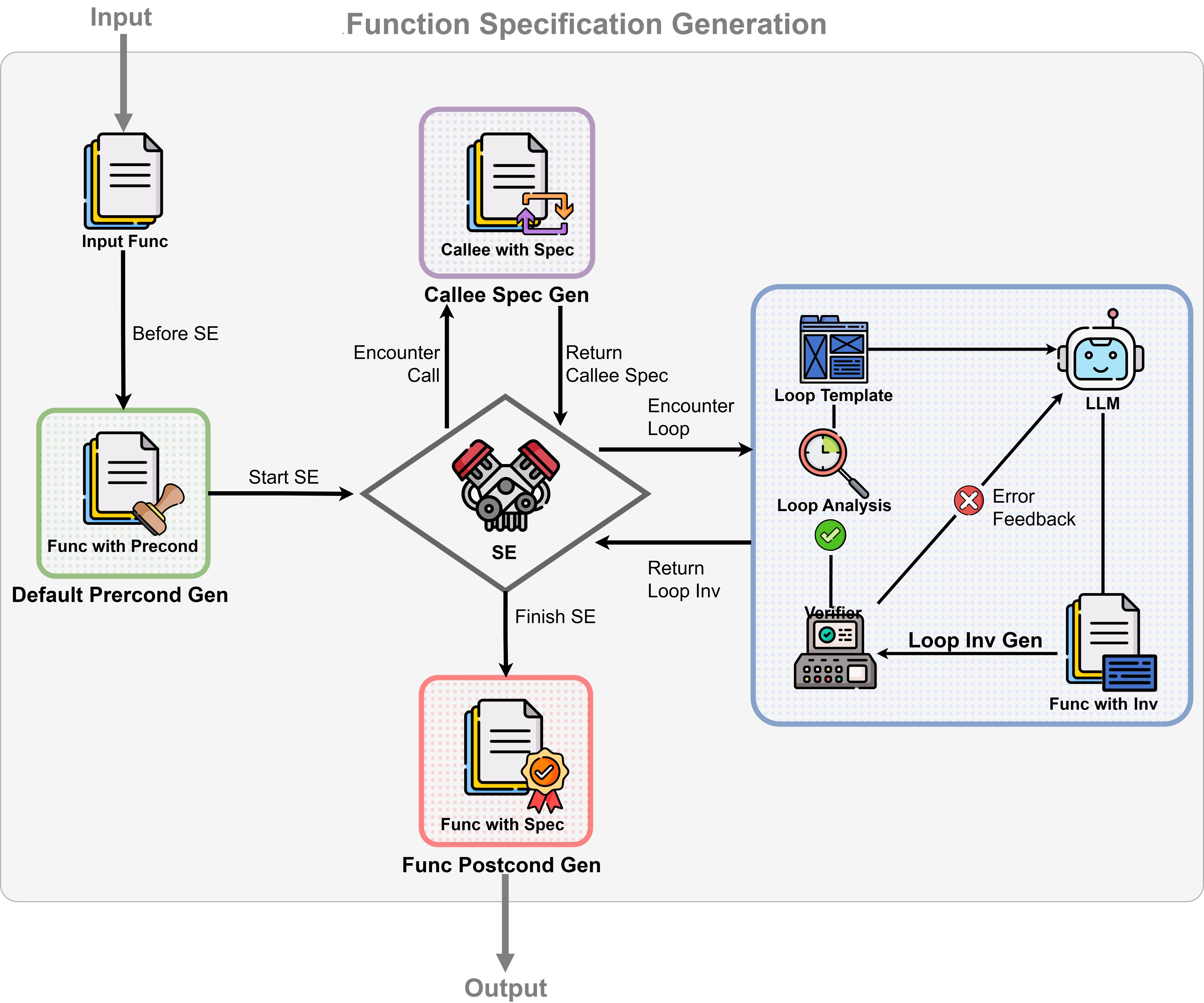}
    \caption{Overview of Our Approach}
    \label{fig:overallachitecture}
\end{figure*}

Fig.~\ref{fig:overallachitecture} shows an overview of our approach.  Without loss of generality, given a function and an optional verification goal, the objective is to automatically generate loop invariants and function pre-/postconditions.  If a verification goal is provided, these specifications must be sufficient to prove it; otherwise, they should be strong enough to capture the function’s core behavior.

The entire process \textbf{Function Specification Generation} comprises four stages, as illustrated in Fig.~\ref{fig:overallachitecture}. It begins with \textbf{Default Precondition Generation} (marked in green), which produces the weakest preconditions ensuring memory-safe execution (node \circled{0} in Fig. \ref{fig:example}). Our tool employs a flattened memory model that recursively decomposes input data structures into atomic memory locations, generating validity and separation constraints to guarantee valid memory accesses. Since other properties beyond memory safety are not required for symbolic execution, they are left for users to specify as needed.

Using this precondition, the tool initiates symbolic execution from the function's entry point (node \circled{1} in Fig. \ref{fig:example}) . When a function call is encountered (node \circled{2} in Fig. \ref{fig:example}), the current symbolic execution is paused and the tool call \textbf{Callee Specification Generation} (mark in purple in Fig. \ref{fig:overallachitecture}). It recursively processes the callee function to generate its specification (node \circled{3} - node \circled{6} in Fig. \ref{fig:example}). Upon completion, the tool returns to the call site, incorporates the generated pre-/postcondition of the callee  and resumes symbolic execution of the caller.
Throughout this process, a call stack is maintained to support nested function calls. Functions are processed in a bottom‑up order,  and all generated function specifications are stored for reuse across different calling contexts.

When symbolic execution encounters a loop within the current function (node \circled{5} in Fig. \ref{fig:example}), it pauses and
transfers control to the \textbf{Loop Invariant Generation} (mark in blue in Fig. \ref{fig:overallachitecture}), the core component of our tool. The Loop Analysis module is first invoked, taking as input the complete symbolic state at the loop entry, and identifies different types of variables by analyzing the behavior of the loop body.
In the presented example, the analysis produces the following results: the unchanged
variables \texttt{pIp->len}, \texttt{pIp} and \texttt{pIp->pkv}; the inductive variables \texttt{i} and \texttt{chksum}. Additionally, the loop entry condition is established, i.e., \texttt{i < pIp->len}. Subsequently, the tool generates parameterized invariant templates designed to capture various execution scenarios, such as whether the loop is entered, which variables remain unchanged,  which evolve inductively and how they behave. These templates are expressed as ACSL  skeletons using equality constraints
to facilitate the synthesis of more precise invariants. Below we present part of the generated templates for the motivation example (see Fig. \ref{fig:example} for the full set):


\[
\small
\begin{aligned}
  & \texttt{loop invariant (0 < \textbackslash at(pIp, Pre)->len) ==> PLACE\_HOLDER\_for\_i;} \\
  & \texttt{loop invariant (0 < \textbackslash at(pIp, Pre)->len) ==> PLACE\_HOLDER\_for\_chksum;} \\ 
  & \texttt{loop invariant pIp == \textbackslash at(pIp, Pre);}
\end{aligned}
\]

These skeletons provide specific \texttt{PLACE\_HOLDER} tags for the LLM to populate (marked as the blue boxes in Fig. ~\ref{fig:example}).  Guided by these structured templates and symbolic execution results, 
LLM produces loop invariants that precisely characterize the behavior of different types of variables.  This structural guidance significantly reduces hallucinations, minimizes syntax errors, and narrows the overall search space. For instance, the generated invariants correctly incorporate the loop-entry condition \texttt{(0 < \textbackslash at(pIp, Pre)->len)}, a critical logical constraint that pure LLM-based approaches consistently overlook.

During the Loop Invariant Generation, a refinement procedure is iteratively invoked to assess the validity (able to pass the formal verification) and the accuracy (sufficient to guarantee the given verification goal) of the generated loop invariants. Depending on the feedback from the verifier, different refinement strategies are applied to guide the LLM toward correct invariants. 

When symbolic execution of the callee is completed (node \circled{6} in Fig. \ref{fig:example}), the \textbf{Function Postcondition Generation} (mark in red in Fig. \ref{fig:overallachitecture}) is triggered, the resulting symbolic states are used to synthesize a postcondition that summarizes the callee's functional behavior. Upon completion, the generated pre-/postconditions are returned to the call site and reused as a functional summary in the caller (node \circled{7} in Fig. \ref{fig:example}), enabling symbolic execution of the caller to resume without re-analyzing the callee's internal control flow.


If symbolic execution has not completed, it proceeds until the final postcondition of the function is generated. Consequently, our approach synthesizes complete functional summaries for this example, including the  weakest preconditions (marked in green) that ensure memory safety, the loop invariants that capture loop-entry conditions, inductive effects, and the identification of unchanged variables, and the postconditions sufficient to prove the given verification goal. Notably, the same postcondition for \texttt{CheckCalFun} can still be generated even when the verification goal is masked, demonstrating that our specification generation is not solely driven by the goal itself.



\section{Pre-/Postcondition Specification Generation}
\label{sec:summarygeneration}
This section first introduces the generation of strongest specifications for code segments without loops via symbolic execution, and then presents the generation of general function specifications. Loop invariant generation is deferred to next section.

\subsection{Symbolic  Execution and Strongest Pre-/Postconditions} %
We use a \emph{flattened memory model} to  explicitly track memory values and aliasing during symbolic execution. In this model, each program point $p$ is associated with a scope consisting of a set of   variables accessible at that point, and the memory at $p$ is modeled as a collection of atomic memory locations $\mathcal{L}_{p}$ reachable from these variables, each of which is treated as a symbolic variable corresponding to a primitive memory unit. 



Within this model, the symbolic execution of statements maintains a symbolic state $(PC_\pi, \sigma_\pi)$ for each execution path $\pi$, where
\begin{itemize}
    \item $PC_\pi$ denotes the path condition, i.e.\ a logical formula over initial symbolic values of $\mathcal{L}_p$ and constants, that expresses the conditions required to traverse path $\pi$. $PC_\pi$ is updated when a branch is taken along $\pi$.
    \item $\sigma_\pi$ denotes the symbolic store along path $\pi$, i.e.\ a logical formula specifying the symbolic values of $\mathcal{L}_p$. $\sigma_\pi$ is updated when statements such as assignments or memory updates execute along $\pi$.
\end{itemize}

The symbolic execution of a program is defined as a transformation over sets of symbolic states, accounting for divergent control flow at branches. Formally, execution begins at the function entry point $p_0$ with an initial symbolic state set containing a single element $(PC_0, \sigma_0)$. Here, $\sigma_0$ assigns initial symbolic values to all variables in the logical variable set $\mathcal{L}_p$, and $PC_0$ is the path condition that encodes the precondition constraining those initial values. For any program point $p$, the set of all reachable symbolic states is given by the collection of pairs $\{(PC_{\pi_i}, \sigma_{p,\pi_i})\}_{i=1}^m$ for every path $\pi_i$ from the entry point $p_0$ to $p$. From this set, we derive a logical formula characterizing all feasible states at $p$, defined as $ \bigvee_{i=1}^{m} (PC_{\pi_i} \land \sigma_{p, \pi_i})$. At the entry point $p_0$, this formula simplifies to the precondition $PC_0 \wedge\sigma_0$. For any subsequent program point, the corresponding formula represents the postcondition that holds up to that point in execution.

We implement a symbolic execution procedure $\textsf{SEStart}(\mathcal{F}, \mathit{Pre}, p_{\mathit{start}})$. Given a code segment $\mathcal{F}$ and a precondition $\mathit{Pre}$ (a formula representing symbolic states) at a starting point $p_{\mathit{start}}$, the procedure executes symbolically from $p_{\mathit{start}}$, exploring the execution paths until it encounters a loop or a function call.


Based on the above definition of symbolic execution, we can derive the \emph{strongest specifications} for loop-free, straight-line code. It consists of a weakest precondition and a strongest postcondition, together giving the most precise description of the input/output behavior.
In our approach, a default weakest precondition is automatically generated, requiring that all accessible atomic memory locations at the entry point are initialized and contain valid symbolic values. Given a precondition $\textit{Pre}$ at the entry  point $p_{\textit{start}}$, the strongest postcondition at  $p_{\textit{end}}$ for $\mathcal{F}$, denoted by $ \textit{SP}(\mathcal{F}, \textit{Pre}, p_{\textit{start}}, p_{\textit{end}})$,
is  the logical formula describing the disjunction of all symbolic states (defined as above) reachable at $p_{\textit{end}}$ from $p_{\textit{start}}$ under $\mathit{Pre}$.

\subsection{Generation of Program Specifications}
\label{sec:functionsummary}


Without loss of generality, we describe the process of generating specifications of functions based on symbolic execution by Algorithm~\ref{alg:funspec-gen}. The algorithm takes as inputs a function $\mathcal{F}$,  a stack $cS$  of callee functions with empty specifications, a list  $aF$ of functions annotated with generated specifications. It performs symbolic execution throughout the entire procedure, recursively invoking the generation of specifications for callee functions and of loop invariants as needed. Upon termination, it produces an annotated version of $\mathcal{F}$ equipped with its specification, including the pre-/postcondition and loop invariants.

\begin{algorithm}[htbp] 
\caption{Function Specification Generation $\mathrm{FuncSpec}$} 
\label{alg:funspec-gen}
\small
\KwIn{$\mathcal{F},\,$callStack $\textit{cS} ,\, $annoFuncs $\textit{aF}$}
\KwOut{Function $\mathcal{F}$ with specification $\mathcal{S}$}


\small
\If{$\mathcal{F} \in \textit{aF}$}{
    \Return
}

$\textit{cS.push}(\mathcal{F})$\;

$\textit{lS} \gets \mathrm{InitLoopStack}()$\;

$\textit{P} \gets \mathrm{DefaultPre}(\mathcal{F})$\;
$ p_{\textit{start}} \gets p_{0} $\;

$\mathrm{SEStart}(\mathcal{F},\textit{P}, p_\textit{start})$\;

\While{$p \neq p_{-1}$}{

$P \gets \textit{SP}(\mathcal{F},\textit{P},p_{\textit{start}},p)$\;
$p_{\textit{start}} \gets p$\;
\If{$p = p_{\textit{loop}}$}{
$\mathcal{F} \gets$
$\mathrm{LoopInvGen}(\mathcal{F},\, p,\, \textit{lS},\,\textit{cS},$
$\,\textit{aF},\, P ,\, \textit{LLM}$)\;
}
 \If{$p = p_{\textit{callee}}$}{
$\mathcal{F}_{\textit{callee}}\gets\mathrm{FuncSpec}(\mathcal{F}_{\textit{callee}},\,\textit{cS},\,\textit{aF})$\;
}
$\mathrm{SEStart}(\mathcal{F},\textit{P}, p_{\textit{start}})$\;
}


$\mathcal{S} \gets \textit{SP}(\mathcal{F},\textit{P},p_{\textit{start}},p_{-1})$ \;

$\mathcal{F} \gets (\mathcal{F},\mathcal{S})$\;

$\textit{aF.append}(\textit{cS.pop()})$\;
\Return $\mathcal{F}$\;
\end{algorithm}

Algorithm \ref{alg:funspec-gen} is explained as follows. 
If $\mathcal{F}$ is already in $aF$, it returns directly (Lines 1-2). Otherwise, it is added to the call stack $cS$, which stores functions encountered during execution that lack specifications,  and a loop stack $ls$ is initialized (Lines 3-4). They  are used to handle nested function calls and loops respectively.  
Lines 5-6 initialize the default precondition  at the entry point $p_0$ (denoted by $P$) and sets the starting point for symbolic execution to be $p_0$ (denoted by $p_{\textit{start}}$).  Function $\mathrm{DefaultPre}$ assigns initial symbolic values to all atomic memory locations reachable from variables in scope at $p_0$ (typically  the function's parameters). 
Symbolic execution of $\mathcal{F}$ then starts from $p_{\textit{start}}$ with precondition $P$ by invoking $\mathrm{SEStart}$  (Line 7), which will stop whenever a loop or a function call is encountered. If the symbolic execution pauses at $p$, which is not the exit point of $\mathcal{F}$ (denoted by $p_{-1}$), the postcondition  $P$ at $p$ is obtained via  $\textit{SP}$ based on the symbolic execution results, and   $p_{\textit{start}}$ (the starting point for resuming symbolic execution in the future) is reset to $p$ (Lines 9-10). If  it is a loop (Lines 11-12), $\mathrm{LoopInvGen}$ is invoked  to synthesize a loop invariant, with $P$ as the loop precondition and $p$ as the loop entry,  and returns the updated $\mathcal{F}$ that annotates the loop with its invariant. If it is a function call (Lines 13-14),  
$\mathrm{FuncSpec}$  is recursively invoked to obtain the specification of the callee function. The use of call stacks enables a  compositional approach to handle nested function calls by first deriving, then leveraging these generated function specifications  across different calling contexts. For both of the cases, the symbolic execution will re-proceed from $p_{\textit{start}}$ with the corresponding annotations (Line 15). 
When the symbolic execution reaches the function's exit, the algorithm derives the postcondition $\mathcal{S}$ of $\mathcal{F}$,   adds the annotated $\mathcal{F}$ with its generated specifications to $\textit{aF}$,  and finally returns  (Lines 16-19).

\section{Loop Invariant Generation} 
\label{sec:loopinvariant}

The general loop invariant generation problem can be defined as follows:
\begin{definition} Given a loop $l:\!$ \texttt{while B do C} of program $\mathcal{P}$,  a precondition $P$ that holds at entry $l$, and an (optional) postcondition $Q$,  the loop invariant generation problem is to find a logical formula $\textit{Inv}$ 
that  satisfies the following conditions:
\begin{itemize}
    \item \textbf{Base:} $P \Rightarrow \textit{Inv}$, i.e., the invariant $\textit{Inv}$ holds initially;
    \item \textbf{Preservation:} $\{\textit{Inv} \land B\} \ C \ \{\textit{Inv}\}$, i.e. if $\textit{Inv}$ holds at the start of a loop iteration and loop condition $B$ is true, executing loop body $C$ preserves $\textit{Inv}$;
\item  \textbf{Termination:} $\textit{Inv} \land \neg B \Rightarrow Q$, i.e. upon loop termination,  $B$ turns false, $\textit{Inv}$ remains true and implies postcondition $Q$.     
\end{itemize}
  \label{Def:invariant}   
\end{definition}

Among these conditions, the postcondition serves as the verification target, whose objective is to synthesize sufficiently strong invariants that imply it. Our invariant generation algorithm, presented below,  is designed to produce strong invariants even in the absence of postconditions. We achieve this through a combination of symbolic execution and Large Language Models (LLMs), using equality-based invariant templates to precisely capture relationships between symbolic states. 
The effectiveness of this approach is experimentally demonstrated in Section~\ref{sec:inv_experiments}.

\begin{algorithm}[htbp] 
\caption{Loop Invariant Generation $\mathrm{LoopInvGen}$} 
\label{alg:loopinv-gen}
\small
\KwIn{$\mathcal{F},\,p_{\textit{loop}} ,\, $loopStack $\textit{lS},\, $callStack $\textit{cS},\,$annoFuncs $\textit{aF},\,$ loop precondition $ P,\,\textit{LLM}$}
\KwOut{Function $\mathcal{F}$ annotated with loop invariant $\mathcal{I}$ for loop $\mathcal{F}_{\textit{loop}}$}


\small
\tcp{\small \textit{$\mathcal{F}_{\textit{loop}}$ is loop body for loop at $p_{\textit{loop}}$ in $\mathcal{F}$}}
$\mathrm{ThinkInNaturalLanguage}(\mathcal{F}_{\textit{loop}},LLM)$\;
$\textit{lS}.push(\mathcal{F}_{\textit{loop}})$\;
$(\mathcal{F}_{\textit{loop}},P)$\;
$\mathcal{F}_{loop}, \textit{uV}, \textit{nV} \gets \mathrm{LoopAnalysis}(\mathcal{F}_{\textit{loop}},P,\textit{lS},\textit{cS},$
$\textit{aF},\textit{LLM})$\;

$\textit{lS}.pop()$\;
$\mathrm{Assert}(\textit{lS}.empty())$\;

$\mathcal{T} \gets $
$\mathrm{LoopInvTemGen}(\mathcal{F}_{\textit{loop}},P,\textit{uV},\textit{nV})$\;

$\mathcal{I} \gets \mathrm{OuterLoopLLMCall}(\mathcal{F},\mathcal{F}_{\textit{loop}},P,\mathcal{T},\textit{LLM})$\;

$\mathcal{F} \gets (\mathcal{F},\mathcal{I})$ \;

\While{$\neg \mathrm{Verify}(\mathcal{F})$}{
    $\mathcal{F} \gets \mathrm{LoopInvRefinement}(\mathcal{F},\textit{LLM})$\;
}

\Return $\mathcal{F}$\;
\end{algorithm}

\subsection{Invariant Generation for Loops}
Algorithm~\ref{alg:loopinv-gen} presents a procedure for the invariant generation 
of a loop in $\mathcal{F}$. Among the inputs, $p_{\textit{loop}}$ denotes the entry point of the loop in $\mathcal{F}$, $P$ is the precondition before $p_{\textit{loop}}$ obtained through previous symbolic execution, the large language model (LLM) serves as the reasoning component. Other parameters remain as previously defined.

First of all (Line 1),  the original outermost $\mathcal{F}_{\textit{loop}}$ is sent to LLM for chain-of-thought in natural language, where LLM were asked few questions for comprehensive understanding. $\mathcal{F}_{\textit{loop}}$ is pushed into the loop stack and analyzed by $\mathrm{LoopAnalysis}$,  which performs the symbolic execution of the loop body to obtain the set of unchanged and non-inductive variables  and furthermore recursively handle all inner loops and annotates their invariants into $F_{loop}$, and after that, the outermost loop $\mathcal{F}_{\textit{loop}}$ is popped from the stack. Third (Lines 6-8), the invariant for the outermost loop is generated: The results obtained from  $\mathrm{LoopAnalysis}$ serve as inputs to $\mathrm{LoopInvTemGen}$, which generates a set of invariant templates corresponding to the current  loop. Based on these invariant templates, the invariant synthesis for the outermost loop is performed by invoking the LLM-based procedure $\mathrm{OuterLoopLLMGen}$. 
At the end (Lines 9-11), $\mathrm{LoopInvRefinement}$ carries out verification-driven refinement, 
where the LLM interacts iteratively with the verifier to repair and strengthen the generated invariants 
until they are verified.

\begin{algorithm}[htbp] 
\caption{loop Analysis $\mathrm{LoopAnalysis}$}  
\label{alg:loop-analysis}
\small
\KwIn{$\mathcal{F},\,$precondition $\textit{Pre},\,$loopStack $\textit{lS} ,\, $callStack $\textit{cS} ,\, $ annoFuncs $\textit{aF}$,\, \textit{LLM}}
\KwOut{$\mathcal{F},$ unchanged and noninductive variables $ \textit{uV},$  $\textit{nV}$}


\small

$\mathcal{F_{\textit{unfolded}}} \gets \mathrm{UnfoldLoop}(\mathcal{F}_{loop})$\;
$P \gets \textit{Pre}$\;
$p_{\textit{start}} \gets p_{\textit{unfolded}\_0}$\;
$\mathrm{SEStart}(\mathcal{F}_{\textit{unfolded}},\textit{P}, p_\textit{start})$\;

\While{$p \neq p_{\textit{unfolded}\_-1}$}{
... \tcp{\small \textit{Same to $\mathrm{FuncSpec}$, reset new precondition and starting point}}
    \If{$p = p_{\textit{nestedLoopEntry}}$}{
$\textit{lS}.push(\mathcal{F}_{\textit{nested}})$\;}

\If{$p = p_{\textit{nestedLoopExit}}$}{
$\mathcal{F}_{\textit{nested}} \gets \textit{lS}.pop()$\;
$\mathcal{F} \gets $
$\mathrm{NestedLoopLLMCall}(\mathcal{F},\mathcal{F}_{\textit{nested}},\textit{LLM})$\;  
}
\If{$p = p_{\textit{callee}}$}{
$\mathcal{F}_{\textit{callee}} \gets$
$\mathrm{FuncSpec}(\mathcal{F}_{\textit{callee}},\,\textit{cS},\,\textit{aF})$\;
}
$\mathrm{SEStart}(\mathcal{F}_{\textit{unfolded}},\textit{P}, p_\textit{start})$\;
}


... \tcp{\small \textit{Same to $\mathrm{FuncSpec}$,  construct the final postcondition   $\mathcal{S}$}}

$\textit{uV} \gets \mathrm{FetchUnchangeVars}(\mathcal{S})$\;

$\textit{nV} \gets \mathrm{FetchNonInductiveVars}(\mathcal{S})$\;

\Return $\mathcal{F},\textit{uV},\textit{nV}$ \;
\end{algorithm}

\subsection{Loop Analysis}
Algorithm~\ref{alg:loop-analysis} implements $\mathrm{LoopAnalysis}$, which generates invariants for nested loops in $\mathcal{F}_{\textit{loop}}$ and identifies its unchanged and non-inductive variables. It performs symbolic execution on a single-iteration unfolding $\mathcal{F}_\textit{unfolded}$ (Lines 1-4).
The key difference with $\mathrm{FuncSpec}$ lies in the handling of nested loops: whereas outer loops use symbolic execution-derived loop preconditions and templates to guide LLM-based invariant generation ($\mathrm{OuterLoopLLMCall}$), nested loops are processed purely via LLM ($\mathrm{NestedLoopLLMCall}$), which infers invariants directly from the code context without relying on preconditions and templates, due to the lack of well-defined preconditions of nested loops. As shown in Algorithm~\ref{alg:loop-analysis}, nested loops are processed in a bottom-up manner using a loop stack $\textit{lS}$, while function calls are handled consistently with $\mathrm{FuncSpec}$ with the help of a call stack (Lines 7-13). Upon completion of the symbolic execution of the outermost loop body, the final postcondition is derived, and all nested loops are annotated with the generated invariants; This postcondition is then used to identify unchanged and non-inductive variables within the outermost loop (Lines 15-17). 


\subsection{Template Generation}
\label{alg:template}


The $\mathrm{LoopInvTemGen}$ procedure takes as inputs the loop body, the loop precondition and the sets of unchanged and non-inductive variables,  and generates a set of invariant templates. These templates have parameters partially determined by symbolic execution results, while the remaining undetermined parameters are completed using the subsequent LLM synthesis method (i.e. $\mathrm{OuterLoopLLMCall}$ in Algorithm~\ref{alg:loopinv-gen}). 
Below we introduce five forms of invariant templates.

\begin{itemize}
    \item Unchanged variables:
For any variable $v \in uV$ and any execution path $\pi_j$ to the loop entry $p$, $v$  retains its symbolic value $\sigma_{p, \pi_j}(v)$ at the loop entry:
\[ PC_{\pi_j} \Rightarrow v = \sigma_{p, \pi_j}(v) \]

\item  Loop skip conditional property: 
For a path $\pi_j$ to $p$ and a loop condition $B$, if $B$ for path $\pi_j$ is false, the loop is skipped without altering variables:  
\[\neg (B \land (PC_{\pi_j} \land \mathrm{\sigma_{p, \pi_j}})) \Rightarrow v = \sigma_{p, \pi_j}(v)\]

\item Non-inductive variables: 
For a non-inductive variable $v$ (whose value is updated independently of its value  from the previous iteration) and any path $\pi_j$ to $p$:
\[(B \land (PC_{\pi_j} \land \mathrm{\sigma_{p, \pi_j}})) \Rightarrow (v = \sigma_{p, \pi_j}(v)) \lor \Phi(v)\]
where $\Phi(v)$ is the inductive invariant to be synthesized.

\item  Inductive variables: 
For an inductive variable $v$ (which is neither unchanged nor non-inductive) and any path $\pi_j$ to $p$:
\[ (B \land (PC_{\pi_j} \land \mathrm{\sigma_{p, \pi_j}})) \Rightarrow \Phi(v) \]

\item Verification goals: 
In addition to variable-specific templates, we also introduce templates corresponding to each verification goal. 
For an verification goal $g$ and any path $\pi_j$ to $p$:
\[ (B \land (PC_{\pi_j} \land \mathrm{\sigma_{p, \pi_j}})) \Rightarrow \Phi(g)
\]

\end{itemize}

This template-driven approach is crucial for our loop invariant generation algorithm. By leveraging precise information derived from symbolic execution, it simplifies the LLM's task, shifting from full invariant generation from scratch to completing  invariant templates. It also  offers structural guidance, helping the LLM produce formulas that conform to the required ACSL logical syntax. More importantly, this approach ensures the generation of \emph{strong invariants} that capture all relevant variable behaviors across every execution path, thereby closely over-approximating the set of all possible variable updates within the loop.

\subsection{Loop Invariant Refinement}
Loop invariants generated by the LLM are not always valid (able to pass formal verification) or sufficiently strong to guarantee the verification goal. 
Also, our template-based approach improves the precision of invariant generation, 
but also constrains the diversity of invariants. 
To alleviate these,  we propose a fine-grained iterative refinement framework, as shown in Algorithm~\ref{alg:inv-repair}, to combine verifier feedback with the LLM’s self-correction ability. This framework is motivated by the observation that LLMs often produce loop invariants that are inaccurate, and that the templates, while improving precision, reduce diversity.

\begin{algorithm}[htbp]
\caption{Iterative Refinement $\mathrm{LoopInvRefinement}$}
\label{alg:inv-repair}
\small
\KwIn{$\mathcal{F}$ with candidate loop invariants $\mathcal{I}_{0}$, \textit{LLM}}
\KwOut{$\mathcal{F}^*$ with verified invariants $\mathcal{I}^*$}

$\mathcal{I} \gets \mathcal{I}_{0}$\;

$\mathcal{F}^{-} \gets \mathcal{F} - {I}_{0}$

 \tcp{\small \textit{Run verifier and collect errors}
 }
 
\While{ $\mathit{Errors} \gets \neg \mathrm{Verify}(\mathcal{F}) $}{

      \If{$\textit{reach iteration limit}$}{
      \tcp{\small \textit{Find largest valid subset in $\mathcal{I}$}}
     \While{$\mathcal{I}$ remains invalid} {
         \ForEach{$ e \in \mathit{Errors} $}{
          $i\gets \textit{associate loop invariant with e}$\;
             $\mathcal{I} \gets \mathrm{Elimination}(\mathcal{I},i)$\;
         }
        
    }
    $\mathcal{F} \gets (\mathcal{F}^{-} ,\mathcal{I})$\;
    $\textbf{break}$\;
    }

     \uIf{$\mathit{Errors} = \textit{syntax error}$}{
        $\mathcal{I} \gets \mathrm{RepairCallLLM}(\mathcal{I})$\;  
    
    }\Else{

        
    

    $\mathit{guidance} \gets \{ \}$ \;

        \uIf{$Errors = \textit{verification goals fail} $ }{
        \tcp{\textit{Only termination failed}}
        $\mathit{guidance} \gets \mathit{guidance} \cup \mathrm{strengthGuide}(\mathcal{I})$\;
        }
        \Else{

        \ForEach{$ e \in \mathit{Errors} $}{
        \uIf{$ e = \textit{loop invariant error}$}{
        $i\gets \textit{associate loop invariant with e}$\;
        \uIf{\textit{e not base}}{
            $\mathit{guidance} \gets \mathit{guidance} \cup \mathrm{WeakenGuide}(i)$\;
        }
        \uIf{\textit{e not preservation}}{
             $\mathit{guidance} \gets \mathit{guidance} \cup \mathrm{AdjustGuide}(i)$\;
        }
        }\Else{
              $\mathit{guidance} \gets \mathit{guidance} \cup \mathrm{RegenGuide}(i)$\;
        }
        }
    }
    }

    $\mathcal{I} \gets\mathrm{RefineCallLLM}(\mathcal{I},\textit{guidance})$\;  
    $\mathcal{F} \gets (\mathcal{F}^{-} ,\mathcal{I})$\;
    
    }

\Return{$\mathcal{F}$}
\end{algorithm}

During the refinement process, depending on the verifier’s feedback, different strategies are applied: 
Syntax errors are repaired directly, 
weak invariants (passing Base and Preservation but not Termination) are strengthened, 
invariants failing Base or Preservation but still implying the Postcondition are weakened or adjusted, 
and invariants failing all checks are discarded and regenerated. 
At each step, the LLM is guided by a specialized prompt that encodes these refinement strategies. If repeated attempts fail, a final elimination step removes consistently invalid invariants.
The refinement continues until a correct invariant is achieved or an iteration limit is reached. Similarly, the refinement of a function specification can apply these strategies.


\oomit{Candidate invariants $\mathcal{I}_0$ are first proposed by the LLM and verified. 
If errors are detected, refinement is triggered. 
The algorithm first prunes invalid candidates by isolating the largest valid subset $\mathcal{I}' \subseteq \mathcal{I}$. 
If $\mathcal{I}'$ already passes verification, the process terminates; 
otherwise refinement continues on the failing invariants. }



\section{Implementation}
\subsection{Toolchain Integration and Data Flow}

\begin{figure*}[h]
    \centering
    \includegraphics[width=0.8\textwidth]{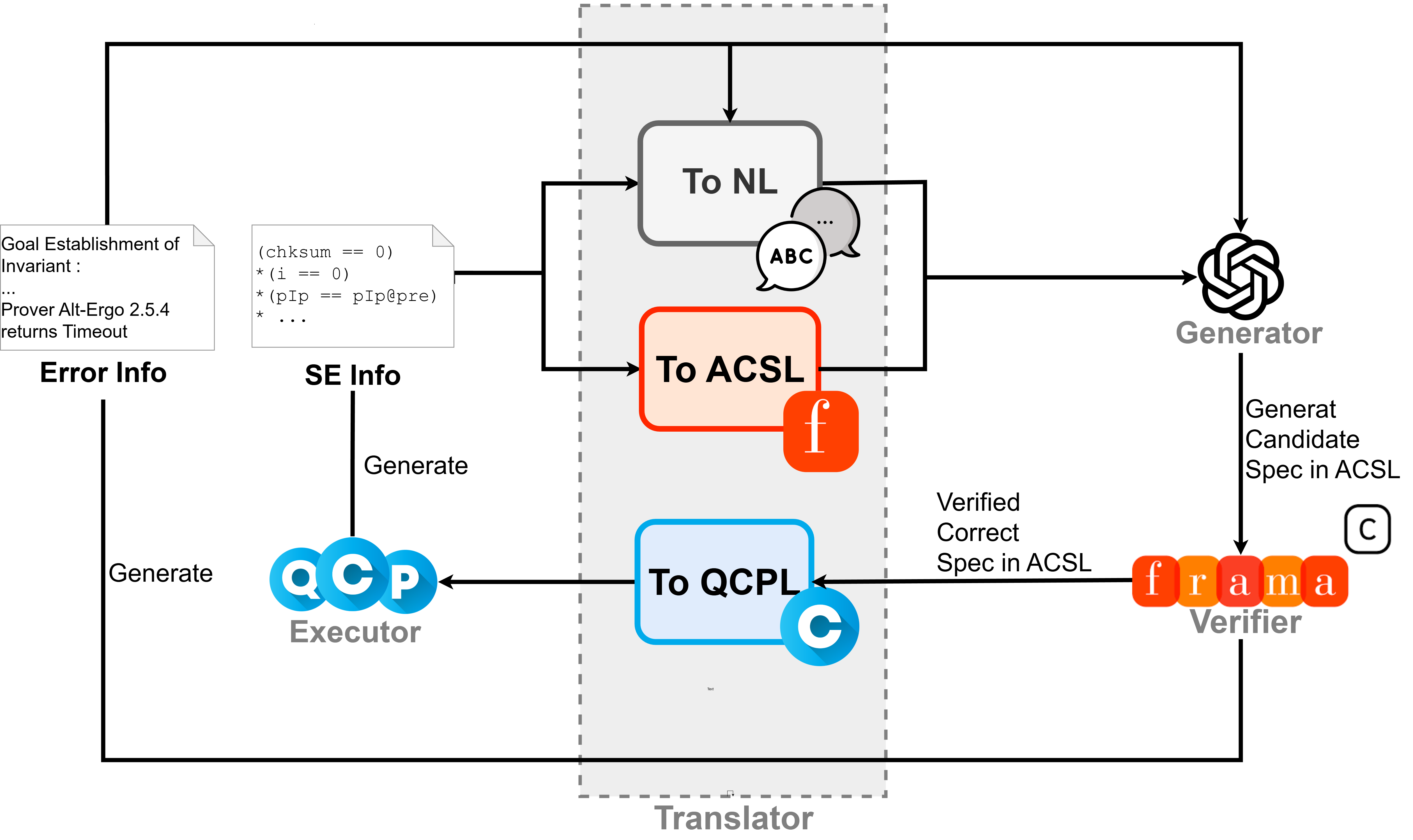}
    \caption{The Implementation Design Framework}
    \label{fig:impl}
\end{figure*}

\begin{figure*}[h]
    \centering
    \includegraphics[width=1.0\textwidth]{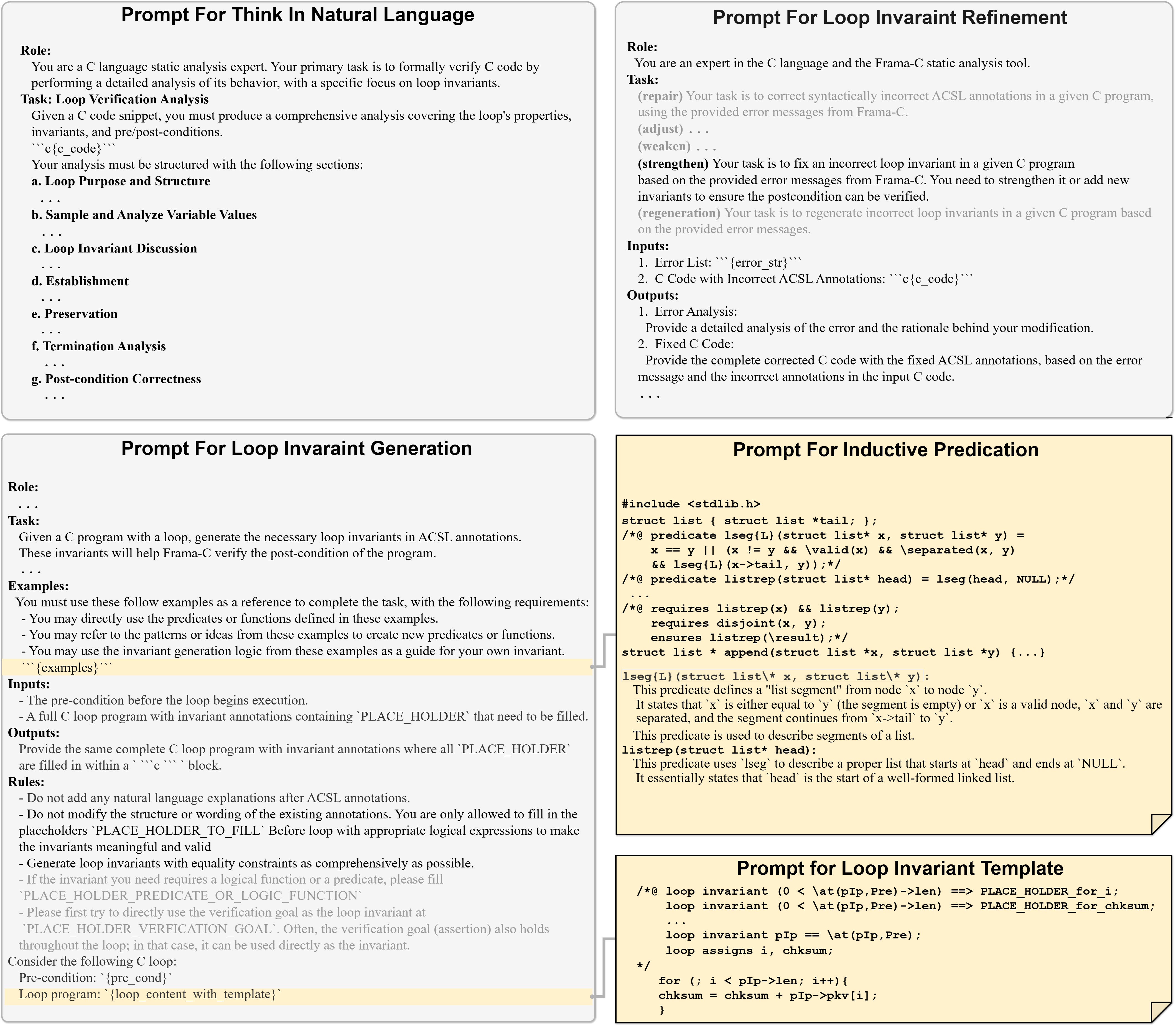}
    \caption{The Prompt Design}
    \label{fig:prompt}
\end{figure*}

The implementation architecture and the underlying data flow between the components are illustrated in Fig.~\ref{fig:impl}.
The process initiates with the C source code entering the \textbf{Executor}, which performs symbolic execution to generate \textit{SE Info} in the form of symbolic states (pairs of path conditions and symbolic stores). This symbolic information is then passed to the \textbf{Translator}, which transforms the symbolic states in the syntax of the  \textbf{Executor} into Natural Language descriptions and the input syntax of the formal \textbf{Verifier}. Based on these translated representations, the \textbf{Generator} synthesizes candidate  specifications, which are subsequently sent to the \textbf{Verifier} for automated verification. If a specification is successfully verified, it is translated back and fed into the \textbf{Executor} for further symbolic execution, allowing the system to progressively cover the entire program. If verification fails, the resulting \textit{Error Info} is translated into Natural Language together with the original error context and fed back to the \textbf{Generator} for iterative refinement. This feedback-driven loop tightly couples symbolic execution, specification synthesis, and formal verification, ensuring that the final output specification is both semantically meaningful and formally sound.

This architecture is instantiated using QCP~\cite{wu2025qcppracticalseparationlogicbased,vst-a}  as the symbolic execution engine and Frama-C~\cite{cuoq2012frama} as the formal verifier, a combination motivated by their complementary strengths. QCP serves as our foundational symbolic executor because it generates Coq-verified postconditions and exposes symbolic execution results at arbitrary control points in a logical form. Moreover, QCP provides strong support for programs manipulating complex data structures, along with a rich library of verified inductive predicates and lemmas, which aligns well with our goal of generating formally precise specifications. In contrast, widely used symbolic execution tools such as Klee~\cite{DBLP:conf/osdi/CadarDE08} neither expose intermediate symbolic states at arbitrary program locations nor represent execution states in a logic-oriented form suitable for direct specification synthesis. However, QCP requires manual annotations for loop invariants and callee function pre/postconditions, and may introduce unproven lemmas that require additional Coq proofs. To mitigate these limitations, we leverage the \textbf{Generator} to automatically synthesize candidate invariants, and integrate Frama-C as the \textbf{Verifier} to discharge proof obligations automatically. Frama-C’s ACSL assertion language provides an LLM-friendly interface for specification generation, while its integration with SMT solvers enables scalable and automated verification. Together, QCP and Frama-C form a synergistic execution–verification backbone that supports our iterative, LLM-in-the-loop specification generation workflow.

The \textbf{Translator} serves as a critical interoperability layer in our system, mediating the interaction between the \textbf{Executor} (QCP), the \textbf{Generator}, and the \textbf{Verifier} (Frama-C). It establishes a bidirectional translation between ACSL and the QCPL (QCP's annotation language), enabling symbolic execution and formal verification to be tightly integrated in an iterative loop. During the $\mathrm{LoopInvGen}$ phase, specifications in ACSL are translated into QCPL so that QCP can interpret them consistently across loop and function boundaries. Conversely, symbolic execution results produced by QCP are translated back into ACSL and discharged by Frama-C through automated verification. In this way, the Translator enables continuous information flow between execution and verification. A key design choice of the \textbf{Translator} is that the internal QCPL representation is never exposed to the LLM. Instead, symbolic execution results are transformed into a combination of Natural Language explanations and ACSL skeletons before being passed to the \textbf{Generator}. This transformation shields the model from low-level syntactic and semantic details specific to QCPL, reducing ambiguity and preventing the model from confusing QCPL with ACSL during generation. At the same time, the use of ACSL skeletons provides a pre-structured formal framework, allowing the LLM to focus on logical correctness rather than grammatical well-formedness.

\subsection{Prompt Designs}

\noindent\paragraph{\textbf{Prompt Design for Loop Invariant Generation}}
Our prompt design strategy illustrated in  Fig.\ref{fig:prompt} is intricately mapped to the multi-stage control flow of Algorithm \ref{alg:loopinv-gen} ($\mathrm{LoopInvGen}$), ensuring that the LLM’s role dynamically adapts to each specific execution phase. The process initiates with the "Think In Natural Language" prompt ($\mathrm{ThinkInNaturalLanguage}$, Line 1) to establish initial program comprehension, followed by the "Loop Invariant Generation" prompt ($\mathrm{OuterLoopLLMCall}$, Line 8), which constrains the LLM to fill precise, symbolic execution-derived Loop Invariant Template, providing a structured skeletal format for ACSL annotations ($\mathrm{LoopInvTemGen}$, Line 7). Crucially, the differential prompting logic is applied across both stages: during the generation phase, prompts and placeholders for arrays or inductive predicates are omitted when irrelevant to the program context; during the refinement procedure ($\mathrm{LoopInvRefinement}$, Line 10), the system dynamically selects appropriate instruction sets—such as repair, weakening, adjustment, strengthening, or regeneration—thereby guiding the LLM to address the specific logical deficiencies identified by the verifier.


\noindent\paragraph{\textbf{Prompt Design for Inductive predication}}
For numerical programs, the specifications can be expressed using arithmetic formulas in first-order logic, and
our symbolic execution-guided LLM approach provides high accuracy for inferring loop invariants.
However, programs involving recursive data structures or inductive operations require specifications expressed using inductive predicates or recursive functions. In these cases, LLMs often struggle not with understanding program semantics, but with correctly adhering to the syntactic and semantic conventions of inductive specifications in the target specification language, frequently producing assertions with syntax errors or uninterpreted functions that lack formal definitions.

To address this issue, in the prompt design (see Fig. ~\ref{fig:prompt}), we provide two textbook-style illustrative examples in Frama-C format, drawn from publicly available and well-established sources: one over arrays and one over single linked lists, taken from the QCP and Frama-C repositories, respectively. These examples are not used to convey program-specific properties, nor to bias the model toward particular benchmark structures. Instead, they serve solely to calibrate the LLM to the proper syntactic forms and intended semantic usage of inductive predicates and functions, thereby reducing hallucinations and malformed specifications.

With this lightweight calibration, the LLM is able to correctly instantiate and synthesize inductive predicates when required, while relying on symbolic execution to infer program-specific behavior. As a result, our approach can handle programs manipulating single linked lists and arithmetic inductive computations (e.g., factorial), which remains beyond the capabilities of~\cite{autospec}—all without resorting to training or large-scale external data.




\section{Experiments}
\label{sec:experiments}

This section evaluates our approach for automatic specification generation based on symbolic execution (hereafter referred to as \toolname).

\subsection{Experimental Setup}
\paragraph{Benchmark}
We evaluate our approach on multiple prominent and widely-adopted benchmarks, which are commonly used by recent state-of-the-art tools. These include:  \textbf{SyGuS}~\cite{alur2019sygus}, collected by CODE2INV~\cite{Code2Inv_a_Deep_Learning_Framework_for_Program_Verification}; \textbf{OOPSLA}~\cite{Inductive_Invariant_Generation_via_Abductive_Inference}; \textbf{NLA}, collected by LIPuS~\cite{Loop_Invariant_Inference_through_SMT_Solving_Enhanced_Reinforcement_Learning}; the datasets  from \textbf{SV-COMP} and \textbf{Frama-C}, collected by AutoSpec~\cite{autospec}.
In addition, we constructed three new benchmarks to   capture  more program scenarios: The \textbf{pIp} set is extracted from an actual Sun Search System used in aerospace control software, involving pointer and struct manipulations as well as complex branching structures. We carefully adapt and reformulate \textbf{pIp} programs into a format acceptable to Frama-C and QCP, thereby establishing a new dataset, which constitutes one of our contributions;
the \textbf{list-S} set is derived from the LIG-MM\cite{LIG-MM} benchmark and includes all singly linked list programs from SLING and SV-COMP; and the \textbf{numer-S} set is a curated collection of additional programs, comprising 20 linear and 20 nonlinear loops primarily from CLAUSE2INV~\cite{DBLP:conf/osdi/CadarDE08}, which exceed the difficulty of those in \textbf{SyGuS} and \textbf{NLA}.


\paragraph{Verifier}

All experiments are performed using Frama-C version 29.0 (Copper). We employ the WP plugin as the verification backend, which enforces a 3-second timeout per query, uses Z3 as the primary SMT solver, and applies the Typed memory model.  

\paragraph{Model}
Our LLM selection for evaluating specification generation balanced model maturity, prevalence, and cutting-edge performance:
\textbf{GPT-3.5-turbo} is an earlier, foundational model that serves as an essential baseline; \textbf{GPT-4o-mini}, a smaller, established model optimized for efficiency; \textbf{GPT-4o}, a widely adopted model for general code comprehension and generation; and \textbf{Claude-3.7-sonnet}, a stronger hybrid model with advanced reasoning and code understanding, representing the top tier of our tool's capabilities. For all experiments, we use a fixed sampling temperature of $0.7$ to balance output diversity and stability.

Below we evaluate  our approach through four research questions:
\begin{itemize}
    \item RQ1: What Types of Programs can SESpec Manage?
    \item RQ2: How does SESpec Perform on Program Specification Generation?
    \item RQ3: Is SESpec Effective?
    \item RQ4: What are the Advantages of SESpec Over Other Work?
\end{itemize}

\subsection{RQ1: What Types of Programs can SESpec Manage?}

\begin{table*}[htbp]
\centering

\caption{Dataset Overview of Program Types}
\resizebox{\textwidth}{!}{%
\begin{tabular}{
    l
    S[table-format=3.0]
    S[table-format=2.1]
    S[table-format=2.1]
    S[table-format=2.1]
    S[table-format=2.1]
    S[table-format=2.1]
    S[table-format=2.1]
    S[table-format=2.1]
    S[table-format=2.1]
    S[table-format=2.1]
    S[table-format=2.1]
}
\toprule
\textbf{Program Type}  & \textbf{SyGus} & \textbf{OOPSLA}   & \textbf{SV-COMP}  & \textbf{NLA}  & \textbf{numer-S}  & \textbf{Frama-C}  & \textbf{pIp}  & \textbf{list-S} & \textbf{\#Program} & \textbf{\#Solved} \\
\midrule
Linear Loop  & {\cellcolor{blue!50}133} & {\cellcolor{blue!35}30} & {\cellcolor{blue!20}14} & {\cellcolor{blue!5}0} & {\cellcolor{blue!25}20} &  {\cellcolor{red!13}7} &   {\cellcolor{red!5}0}  &{\cellcolor{red!5}0} & 204 & 197\\
Non Linear Loop  & {\cellcolor{blue!5}0} & {\cellcolor{blue!5}0} & {\cellcolor{blue!5}0} & {\cellcolor{blue!35}30} & {\cellcolor{blue!25}20} &  {\cellcolor{red!8}3} &  {\cellcolor{red!5}0} & {\cellcolor{red!5}0} & 53 & 41\\
Nested \& Multi Loop  & {\cellcolor{blue!5}0} & {\cellcolor{blue!22}16} & {\cellcolor{blue!12}7} & {\cellcolor{blue!5}0} & {\cellcolor{blue!5}0} &  {\cellcolor{red!6}1} &  {\cellcolor{red!5}0} &{\cellcolor{red!5}0} & 24 & 15 \\
Memory \& Loop  & {\cellcolor{blue!5}0} & {\cellcolor{blue!5}0} & {\cellcolor{blue!7}2} & {\cellcolor{blue!5}0} & {\cellcolor{blue!5}0} &  {\cellcolor{red!20}17}  &  {\cellcolor{red!18}12} & {\cellcolor{red!20}24} & 53 & 45  \\
Iteration Free & {\cellcolor{blue!5}0} & {\cellcolor{blue!5}0} & {\cellcolor{blue!5}0} & {\cellcolor{blue!5}0} & {\cellcolor{blue!5}0} &  {\cellcolor{red!30}20} &  {\cellcolor{red!44}38} & {\cellcolor{red!5}0} & 58& 56 \\
External \&  Recursive  Call & {\cellcolor{blue!5}0} & {\cellcolor{blue!5}0} & {\cellcolor{blue!5}0} & {\cellcolor{blue!5}0} & {\cellcolor{blue!5}0} &  {\cellcolor{red!8}3} &  {\cellcolor{red!5}0} & {\cellcolor{red!5}0} & 3 & 0 \\
\midrule
\textbf{\#Program} & 133 & 46 & 21 & 30 & 40 &  51 & 50 & 24 & 395  \\
\textbf{\#Solved} & 133 & 40 & 16 & 25 & 34 &  39 & 44 & 23 & & 354 \\
\textbf{Avg. LoC} & 19& 24 & 22 &21 & 19& 18 & 57 & 24\\
\bottomrule
\end{tabular}%
}
\label{tab:dataset_overview}
\end{table*}




Table~\ref{tab:dataset_overview} summarizes the 395 benchmark programs used in our evaluation. They encompass a diverse range of program structures, including linear/non-linear/nested/multi-loops, function calls, along with data structures involving pointer manipulations, and recursively-defined data types such as linked lists. Importantly, our benchmark suite includes both loop-only programs (indicated by \textcolor{blue}{Blue} cells), requiring invariant generation, and programs requiring full specification generation, i.e. loop invariant/pre/postcondition (indicated by \textcolor{red}{Red} cells).


Overall, SESpec successfully verified 354 out of 395 programs (nearly 90\%). The results demonstrate that our method is applicable to a wide range of program types, especially generalizes to real-world industrial cases in the \textbf{pIp} dataset selected from an actual Sun Search System. Our current approach does not yet cover recursive functions or programs involving library functions and system calls, which we leave as future work.

Furthermore, our evaluation reveals that lines of code (LOC) do not reliably reflect verification difficulty on these benchmarks; instead, the primary challenge arises from the complexity of loop invariants. As shown in our results, SESpec achieves a 97\% success rate on Linear Loop and Iteration-Free programs. However, the success rate drops to 77\% for Non-Linear Loop programs and to 63\% for Nested \& Multi-Loop programs. This decline is primarily due to the increased difficulty of synthesizing invariants that capture complex numerical relationships, intricate data structure properties, and interactions among nested or multi-level loop structures.

\subsection{RQ2: How does SESpec Perform on Program Specification Generation?}

\begin{table*}[h!] 
\centering
\caption{Specification Generation Performance Overview} 
\label{tab:performance_rates_model} 

\scriptsize
\setlength{\tabcolsep}{2pt}
\renewcommand{\arraystretch}{1.0}
\resizebox{\textwidth}{!}{
\begin{tabular}{cc ccc ccc ccc}
\toprule
\multirow{2}{*}{\textbf{Benchmark}} & \multirow{2}{*}{\textbf{Model}}
& \multicolumn{3}{c}{\textbf{Pass@1 (\%)}} 
& \multicolumn{3}{c}{\textbf{Pass@3 (\%)}} 
& \multicolumn{3}{c}{\textbf{Pass@5 (\%)}} \\
\cmidrule(lr){3-5} \cmidrule(lr){6-8} \cmidrule(lr){9-11}
& & \textbf{Syn.} & \textbf{Val.} & \textbf{Acc.}
& \textbf{Syn.} & \textbf{Val.} & \textbf{Acc.}
& \textbf{Syn.} & \textbf{Val.} & \textbf{Acc.} \\
\midrule

\multirow{4}{*}{\textbf{SyGus}}
 & GPT-4o-mini & 90.98 & 87.97 & 78.20 & 97.74 & 95.49 & 88.72 & 98.50 & 96.99 & 90.23 \\
 & GPT-3.5-turbo & 93.23 & 88.72 & 84.21 & 97.74 & 97.74 & 92.48 & 98.50 & 98.50 & 93.23 \\
 & GPT-4o & 98.50 & 98.50 & 93.98 & 100.00 & 99.25 & 99.25 & 100.00 & 100.00 & 100.00 \\
 & Claude-3.7-sonnet & 100.00 & 100.00 & 100.00 & 100.00 & 100.00 & 100.00 & 100.00 & 100.00 & 100.00 \\
\midrule

\multirow{4}{*}{\textbf{OOPSLA}}
 & GPT-4o-mini & 89.37 & 78.95 & 65.41 & 89.13 & 86.96 & 43.48 & 89.13 & 86.96 & 47.83 \\
 & GPT-3.5-turbo & 82.61 & 82.61 & 47.83 & 86.96 & 86.96 & 52.17 & 86.96 & 86.96 & 52.17 \\
 & GPT-4o & 91.30 & 69.57 & 60.87 & 95.65 & 76.09 & 73.91 & 100.00 & 82.61 & 76.08 \\
 & Claude-3.7-sonnet & 93.48 & 69.67 & 63.04 & 97.83 & 80.43 & 76.09 & 97.83 & 93.57 & 86.96 \\
\midrule

\multirow{4}{*}{\textbf{SV-COMP}}
 & GPT-4o-mini & 61.90 & 61.90 & 28.57 & 80.95 & 80.95 & 42.86 & 85.71 & 80.95 & 57.14 \\
 & GPT-3.5-turbo & 95.24 & 90.48 & 66.67 & 95.24 & 90.48 & 66.67 & 95.24 & 90.48 & 66.67 \\
 & GPT-4o & 90.48 & 90.48 & 52.38 & 100.00 & 100.00 & 71.43 & 100.00 & 100.00 & 76.19 \\
 & Claude-3.7-sonnet & 100.00 & 100.00 & 71.43 & 100.00 & 100.00 & 71.43 & 100.00 & 100.00 & 76.19 \\
\midrule

\multirow{4}{*}{\textbf{NLA}}
 & GPT-4o-mini & 90.00 & 83.33 & 76.67 & 90.00 & 83.33 & 76.67 & 90.00 & 83.33 & 76.67 \\
 & GPT-3.5-turbo & 76.67 & 70.00 & 50.00 & 80.00 & 80.00 & 56.67 & 80.00 & 80.00 & 56.67 \\
 & GPT-4o & 90.00 & 86.67 & 80.00 & 93.33 & 93.33 & 83.33 & 93.33 & 93.33 & 83.33 \\
 & Claude-3.7-sonnet & 93.33 & 93.33 & 83.33 & 93.33 & 93.33 & 83.33 & 93.33 & 93.33 & 83.33 \\
\midrule

\multirow{4}{*}{\textbf{numer-S}}
 & GPT-4o-mini & 95.00 & 90.00 & 72.50 & 100.00 & 95.00 & 75.00 & 100.00 & 95.00 & 75.00 \\
 & GPT-3.5-turbo & 92.50 & 85.00 & 62.50 & 95.00 & 90.00 & 65.00 & 95.00 & 92.50 & 67.50 \\
 & GPT-4o & 100.00 & 100.00 & 82.50 & 100.00 & 100.00 & 82.50 & 100.00 & 100.00 & 82.50 \\
 & Claude-3.7-sonnet & 100.00 & 97.50 & 85.00 & 100.00 & 100.00 & 85.00 & 100.00 & 100.00 & 85.00 \\
\midrule

\multirow{4}{*}{\textbf{Frama-C}}
 & GPT-4o-mini & 54.90 & 49.02 & 47.06 & 62.75 & 50.98 & 49.02 & 64.71 & 50.98 & 49.02 \\
 & GPT-3.5-turbo & 77.08 & 60.42 & 52.08 & 83.33 & 62.50 & 54.17 & 83.33 & 62.50 & 54.17 \\
 & GPT-4o & 82.35 & 72.55 & 66.67 & 86.27 & 76.47 & 70.59 & 86.27 & 78.43 & 72.55 \\
 & Claude-3.7-sonnet & 86.27 & 78.43 & 72.55 & 92.16 & 82.35 & 76.47 & 92.16 & 84.31 & 76.47 \\
\midrule

\multirow{4}{*}{\textbf{pIp}}
 & GPT-4o-mini & 90.00 & 87.50 & 65.00 & 92.50 & 92.50 & 72.50 & 92.50 & 92.50 & 75.00 \\
 & GPT-3.5-turbo & 84.00 & 76.00 & 66.00 & 88.00 & 78.00 & 72.00 & 88.00 & 78.00 & 76.00 \\
 & GPT-4o & 94.00 & 78.00 & 64.00 & 94.00 & 78.00 & 76.00 & 94.00 & 78.00 & 78.00 \\
 & Claude-3.7-sonnet & 100.00 & 88.00 & 82.00 & 100.00 & 90.00 & 88.00 & 100.00 & 92.00 & 88.00 \\
\midrule

\multirow{4}{*}{\textbf{list-S}}
 & GPT-4o-mini & 62.50 & 41.67 & 33.33 & 75.00 & 45.83 & 41.67 & 79.17 & 54.17 & 45.83 \\
 & GPT-3.5-turbo & 70.83 & 20.83 & 8.33 & 79.17 & 25.00 & 16.67 & 79.17 & 25.00 & 16.67 \\
 & GPT-4o & 70.83 & 62.50 & 54.17 & 87.50 & 83.33 & 75.00 & 91.67 & 87.50 & 79.17 \\
 & Claude-3.7-sonnet & 95.83 & 83.33 & 75.00 & 100.00 & 100.00 & 95.83 & 100.00 & 100.00 & 95.83 \\
\bottomrule
\end{tabular}}
\end{table*}

We evaluate correctness through three metrics:
\begin{itemize}
\item \textbf{Syntax rate (Syn.):} the percentage of generated specifications that are syntactically correct;
\item \textbf{Validity rate (Val.):} the percentage of generated specifications that are syntactically correct and valid, meaning that they can pass the formal verification of verifiers such as Frama-C. For example, the generated invariants for loops are valid if they satisfy both the \textit{Base} and \textit{Preservation} conditions;
\item \textbf{Accuracy rate (Acc.):} the percentage of generated specifications that are syntactically correct, valid, and also satisfy the verification goal. For example, the generated invariants for loops are accurate if they satisfy the \textit{Base}, \textit{Preservation}, and \textit{Termination} conditions.
\end{itemize}
These metrics are assessed under different Pass@k settings, reflecting the system's ability to attempt multiple generations until the specifications pass
all three metrics.

Table~\ref{tab:performance_rates_model} presents a comprehensive overview of our experimental results.

The observed gaps among different evaluation metrics reveal important characteristics of both benchmark difficulty and model capability. As benchmark complexity increases, the discrepancies between Syn., Val., and Acc. become more pronounced. A similar trend is observed across models: weaker models tend to exhibit larger gaps between these metrics. Notably, Val. does not trivially coincide with Syn. in our results, indicating that the generated specifications are not simplistic invariants (e.g., \texttt{true}) or trivial properties that programs would satisfy by default. Instead, our approach aims to capture complete program semantics and functional correctness, which naturally induces a meaningful distinction between syntactic correctness and semantic validity.

Importantly, our method demonstrates consistent effectiveness across different models and significantly narrows the performance gap between stronger and weaker ones. For example, on benchmarks such as \textbf{SV-COMP}, \textbf{NLA}, and \textbf{numer-S},  GPT-4o and Claude achieve identical performance at higher Pass@k levels (Pass@3 and Pass@5), highlighting the capability of our approach to amplify the effectiveness of comparatively weaker models. Furthermore, the method exhibits rapid convergence under Pass@k evaluation. Performance improvements from Pass@3 to Pass@5 are marginal across several benchmarks, including \textbf{NLA}, \textbf{numer-S}, and \textbf{pIp}, suggesting high resource efficiency by reducing the need for excessive evaluation cycles.

The results also indicate comparatively lower accuracy on the \textbf{OOPSLA}, \textbf{SV-COMP}, and \textbf{Frama-C} benchmarks. For \textbf{OOPSLA} and \textbf{SV-COMP}, this is primarily due to the presence of nested and multiple loops, where symbolic execution often fails to derive sufficiently strong preconditions. In addition, the absence of explicit loop conditions in \textbf{OOPSLA} further complicates invariant inference. In the case of \textbf{Frama-C}, the difficulty mainly stems from complex loop structures involving array manipulations, external function calls, and recursive behaviors, which remain challenging for symbolic execution. Despite these obstacles, our method still achieves notable performance on these challenging benchmarks. Moreover, the consistent improvements from Pass@1 to Pass@5 further underscore the robustness and scalability of our approach in complex verification scenarios.

\subsection{RQ3: Is SESpec Effective?}

To assess the individual contributions of each component in our method, we conducted a detailed ablation study using 100 randomly chosen programs, consisting of 70 loop-related and 30 function-related cases. To ensure a fair comparison under controlled computational budgets, we designed the experiment to maintain consistent LLM token usage across all configurations. Specifically, we capped the refinement process at a maximum of 3 iterations.


We compared four distinct configurations: The \textbf{Only LLM (Pass@3)} baseline uses the LLM for trivial invariant generation without any external guidance. The \textbf{LLM + SE (Pass@3)} variant enhances the LLM with symbolic execution, providing structural templates and program context. \textbf{LLM + SE + REFINE (Pass@1)} incorporates our refinement module for iterative improvement. These configurations are designed to facilitate comparison with the incremental approach employed in our work.

\begin{figure*}[htbp]
    \centering
    \begin{minipage}[b]{0.48\textwidth}
        \centering
        \scriptsize
        \setlength{\tabcolsep}{3.5pt}
        \begin{tabular}{l l S[table-format=3.2] S[table-format=3.2] S[table-format=3.2]}
            \toprule
            \textbf{Method} & \textbf{Syn. (\%)} & \textbf{Val. (\%)} & \textbf{Acc. (\%)} \\
            \midrule
            Only LLM & {\cellcolor{blue!45}97.50} & {\cellcolor{blue!20}37.50} & {\cellcolor{blue!15}28.75} \\
            LLM + SE & {\cellcolor{blue!45}95.71} & {\cellcolor{blue!30}62.86} & {\cellcolor{blue!30}60.00} \\
            LLM + SE + REFINE & {\cellcolor{blue!45}95.71} & {\cellcolor{blue!45}95.71} & {\cellcolor{blue!40}87.21} \\
            \midrule
            Only LLM & {\cellcolor{red!30}66.67} & {\cellcolor{red!15}26.67} & {\cellcolor{red!10}16.67} \\
            LLM + SE & {\cellcolor{red!35}76.67} & {\cellcolor{red!25}50.00} & {\cellcolor{red!20}46.67} \\
            LLM + SE + REFINE& {\cellcolor{red!45}93.33} & {\cellcolor{red!40}86.67} & {\cellcolor{red!40}86.67} \\
            \bottomrule
        \end{tabular}
        \caption{Ablation Study. Blue cells: loop invariant generation. Red cells: program specification generation.}
        \label{fig:ablation_table}
    \end{minipage}%
    \hfill
    \begin{minipage}[b]{0.48\textwidth}
        \centering
        \includegraphics[width=1.0\linewidth]{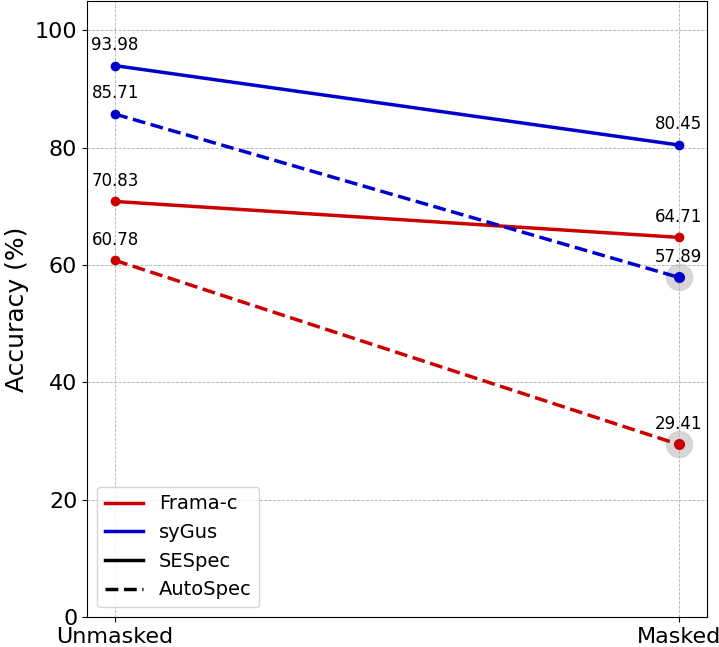}
        \caption{The impact of masking verification
goals for specification generation. Gray-shaded dots are results reproduced by us.}
        \label{fig:mask}
    \end{minipage}
\end{figure*}

The ablation results indicate that using the LLM alone yields high syntactic correctness, but exhibits notably low validity and accuracy. Across both experimental settings, the integration of symbolic execution (SE) leads to substantial improvements in validity and accuracy—with an average increase of 30\%—while only slightly reducing syntactic correctness in the first setting (less than 2\%). The incorporation of the refinement step (REFINE) further elevates performance, achieving perfect scores across all metrics in the first setting and delivering an average improvement of 20\% in validity and accuracy in the second.

Overall, the ablation study confirms that each component contributes positively to the final outcome, with symbolic execution and refinement yielding the most significant gains.



\subsection{RQ4: What are the Advantages of SESpec Over Other Work?}

\begin{table}[htbp]
    \centering
    \caption{Specification Generation Performance Comparison. The {\color{gray!80} \textbf{Gray-shaded}} cells represent results reproduced by us.}
    \label{tab:comparsion}
    \resizebox{\linewidth}{!}{ 
    \begin{tabular}{l cccccccc}
        \toprule
        \textbf{Method} & \textbf{SyGus} & \textbf{OOPSLA} & \textbf{SV-COMP} & \textbf{NLA} & \textbf{numer-S}  & \textbf{Frama-C} &\textbf{pIp} & \textbf{list-S}\\
        \midrule
        \textbf{CODE2INV} & 54.9\% & --- & --- & --- & ---& ---& ---& ---\\
        \textbf{HOLA}   & ---  & \textbf{93.5\%} & --- & --- & 
        --- & ---& ---& ---
        \\
        \textbf{CLN2INV}  & 93.2\% & --- & ---& --- &  --- & ---& ---& ---\\
        \textbf{LiPus} & 93.2\% & --- & --- & 83.3\% & --- &  --- & ---& ---\\
         \textbf{CLAUSE2INV} & 99.2\% & --- & ---& \textbf{93.3\%} &  --- & ---& ---& --- \\
        \midrule
        \textbf{AutoSpec-3.5}  & 85.7\% & 82.6\% & \textbf{76.2\%}& {\cellcolor{gray!20}70.0\% }& {\cellcolor{gray!20}75.0\% } & 60.8\% & {\cellcolor{gray!20}14.0\% } & {\cellcolor{gray!20}0.0\% }\\
        \textbf{AutoSpec-4o}  & {\cellcolor{gray!20}86.5\%} &  {\cellcolor{gray!20}71.7\%} & {\cellcolor{gray!20}66.7\%}& {\cellcolor{gray!20}70.0\% }& {\cellcolor{gray!20}72.5\% } & {\cellcolor{gray!20}68.6\% }& {\cellcolor{gray!20}22.0\% } & {\cellcolor{gray!20}0.0\% }\\
        \textbf{AutoSpec-Claude-3.7}  & {\cellcolor{gray!20}88.7\%} &  {\cellcolor{gray!20}80.4\%} & {\cellcolor{gray!20}66.7\%}& {\cellcolor{gray!20}60.0\% }& {\cellcolor{gray!20}55.0\% } & {\cellcolor{gray!20}74.5\% }& {\cellcolor{gray!20}4.0\% } & {\cellcolor{gray!20}0.0\% }\\
        \textbf{ACInv-3.5} & 71.4\% & 45.7\% & 42.9\% & --- & ---  & 34.5\% & ---  & --- \\
        \textbf{ACInv-4o}  & 79.9\% & 76.1\% & \textbf{76.2\%} & --- & ---  & 48.3\% & ---  & --- \\
        \textbf{SESpec-3.5} & 93.2\% & 52.2\% & 66.7\% & 56.7\% &67.5\% & 51.0\% & 76.0\% & 16.7\%\\
        \textbf{SESpec-4o}     & \textbf{100.0\%} & 76.1\% & \textbf{76.2\%} & 83.3\% & 82.5\% & 72.6\% & 78.0\% & 79.2\% \\
            \textbf{SESpec-Claude-3.7} & \textbf{100.0\%} & 87.0\% & \textbf{76.2\%} &  83.3\% &  \textbf{85.0}\% & \textbf{76.5}\% & \textbf{88.0}\%& \textbf{95.83}\% 
        \\
        \bottomrule
    \end{tabular}
    }
\end{table}

\noindent\paragraph{\textbf{Overall Comparison With LLM Approaches}} 

Table~\ref{tab:comparsion} compares three representative LLM-based approaches: AutoSpec, ACInv, and our tool SESpec.
First, on the four datasets originally introduced by AutoSpec, our method consistently outperforms ACInv across both GPT-3.5-turbo and GPT-4o. Moreover, the performance gains from upgrading the underlying model are similar for ACInv and SESpec, highlighting that our workflow is comparably scalable yet significantly more effective.

In contrast, AutoSpec exhibits negligible improvement when transitioning from GPT-3.5-turbo to GPT-4o. This stagnation stems from AutoSpec’s reliance on few-shot prompting combined with exhaustive sampling and verification-based filtering. By prioritizing the generation of massive candidate pools rather than leveraging the full reasoning potential of LLMs, AutoSpec hits a performance ceiling that stronger models cannot easily break. Consequently, our reproduction of AutoSpec-4o remains at a similar level to the reported AutoSpec-3.5 results.

Conversely, SESpec demonstrates steady and reliable gains as the underlying model improves. Our workflow demands complex reasoning capabilities—specifically, simulating execution via a "Thinking In Natural Language" procedure and iteratively repairing, strengthening, or weakening invariants based on verification feedback. This complexity exceeds the instruction-following limits of GPT-3.5-turbo, resulting in less competitive performance on the weaker model. The use of symbolic execution reduces this complexity by providing structured symbolic constraints,  enabling superior models like GPT-4o to achieve significant performance gains through deeper reasoning, eventually surpassing AutoSpec across almost all benchmarks.

This distinction is further illustrated by the performance on complex datasets like \textbf{pIp}. We observe that AutoSpec-Claude-3.7 performs poorly despite the model's advanced capabilities. Although Claude possesses superior mathematical logic and coding abilities, it often struggles to generate correct ACSL syntax, leading to high failure rates. SESpec mitigates this issue by utilizing an ACSL skeleton to represent symbolic states, which structurally constrains the output to avoid syntactic errors while fully leveraging the model's logical reasoning strengths.

Finally, we emphasize the difference in experimental setup. SESpec operates primarily in a zero-shot setting. In contrast, AutoSpec relies on handcrafted few-shot examples as an indispensable part of its workflow. This highlights the stronger generalization ability of SESpec: it achieves competitive or superior performance through autonomous reasoning rather than depending on few-shot engineering.
\noindent\paragraph{\textbf{Detailed Results over Benchmarks on SESpec}}
Our method demonstrates consistent advantages over SMT-based approaches (combined with machine learning) across several major numerical invariant benchmarks. On the \textbf{SyGuS} dataset introduced by CODE2INV, our approach achieves a 100\% success rate, outperforming the state-of-the-art SMT-based tool CLAUSE2INV. For the nonlinear numerical \textbf{NLA} benchmark proposed by LiPus, our method matches the success rate of LiPus and remains competitive, although it trails CLAUSE2INV by 10\%. Furthermore, on the \textbf{OOPSLA} dataset, proposed by the abductive inference-based invariant generator HOLA and designed for numerical loops without explicit conditions, including nested and multiple loops, our method achieves a success rate of 87\% on this challenging case.
However, these existing approaches based on SMT solvers are specialized for numerical loops. 

By integrating symbolic execution, our method demonstrates significant advantages over purely neural baselines like the state-of-the-art AutoSpec and ACInv on numerical benchmarks. It achieves a 15\% higher accuracy on \textbf{SyGuS} and noticeably outperforms them on \textbf{NLA} and \textbf{numer-S}. On the \textbf{OOPSLA} benchmark, our approach achieves a success rate comparable to ACInv and is only marginally lower than AutoSpec (With the Claude-3.7 model, we achieve 5\% higher rate). All three methods perform equally on \textbf{SV-COMP} benchmark. We speculate that this is due to the dataset's non-uniform distribution of problem difficulty, with challenging nested and multi-loops  that remain unsolved by all existing approaches.
 The above results demonstrate how symbolic guidance enables more precise specification generation that pure neural methods cannot achieve.

On the \textbf{Frama-C} benchmark for functional program verification, our method noticeably outperforms AutoSpec by incorporating support for inductive predicates, enabling more expressive  specification inference. When evaluated on the \textbf{pIp} benchmark from the real-world Sun Search system, which features extensive use of structures and pointer operations, our approach achieves an average 50\% higher success rate compared to AutoSpec,  thanks to the  combination of a flattened memory model and precise symbolic execution. 
Furthermore, our method attains high performance on the \textbf{list-S} benchmark comprising linked lists, a capability absent in AutoSpec due to its lack of support for inductive data types. These results demonstrate that our approach offers robust and generalizable support for both closed struct types (via flattened unfolding) and inductive data structures (through inductive predicates).


\noindent\paragraph{\textbf{Specification Generation without Verification Goals}}To demonstrate the capability of our tool \toolname\ on specification generation in the absence of verification goals, we conducted experiments on \textbf{syGus} and \textbf{Frama-C} by deliberately masking the goals and compared with AutoSpec. We performed GPT-4o with Pass@1 experiments on SESpec and chose their own configuration on AutoSpec~\cite{autospec}, respectively, to compare the accuracy of the two and to analyze the impact of masking verification goals.
As shown in Figure~\ref{fig:mask}, SESpec maintains relatively stable accuracy across the two settings, decreasing from 93.98\% to 80.45\%. Similarly, in the \textbf{Frama-C} benchmark, the accuracy only drops slightly from 70.83\% to 64.71\%, further demonstrating the robustness of SESpec without verification goal. By contrast, AutoSpec experiences a drastic decline in accuracy: On the \textbf{syGus} benchmark, accuracy falls from 85.71\% to 57.89\%, while on \textbf{Frama-C} it decreases sharply from 60.78\% to just 29.41\%.


The performance drop of SESpec is primarily due to the refinement process, which performs targeted optimization with respect to the specific verification goal. A key finding is that if the improvement from refinement is disregarded in our approach, the performance under the two settings is still nearly identical. This demonstrates that our symbolic execution-guided reasoning ensures high-quality specification generation even in the absence of specific verification targets.

\label{sec:inv_experiments}



\section{Threats to Validity}
\label{sec:threats}
\noindent\paragraph{\textbf{Data Leakage}}
A major threat to the validity of specification generation approaches lies in unintended bias, data leakage, and overfitting to evaluation benchmarks, especially in the context of program verification where truly domain-specific data are scarce. Training-based or retrieval-augmented methods are particularly vulnerable to these issues, as benchmark adjacent corpora, curated examples, or synthetically generated programs may implicitly encode benchmark specific patterns, leading to over-optimization for a given dataset rather than genuine semantic understanding.

We mitigate these threats by deliberately avoiding all forms of external supervision and dataset dependency. Our approach does not involve model fine-tuning, retrieval from large-scale external corpora, synthesized datasets, or few-shot prompting. Consequently, the model cannot adapt its behavior to any specific benchmark distribution or overfit to particular datasets. Instead, we exclusively rely on the in-context reasoning capabilities of pretrained large language models, guided by symbolic execution results derived solely from the target program. These symbolic states provide precise semantic context without introducing any external information about benchmark structure or expected specifications.

As a result, the generated specifications are grounded purely in the semantics of the analyzed program rather than in prior exposure to benchmark-related data. This design substantially strengthens the internal validity of our evaluation by minimizing bias, preventing data leakage, and avoiding overfitting to any dataset, ensuring that observed performance gains stem from the proposed symbolic execution–guided reasoning process rather than dataset-specific artifacts.

\noindent\paragraph{\textbf{Benchmark and Baseline Selection}}
To ensure a fair and comprehensive evaluation, we selected benchmarks that align with the current state-of-the-art in program verification. Our evaluation includes benchmarks used in the works we compare against, specifically the AutoSpec benchmark and two external datasets from Lipus and Clause2Inv. These benchmarks are representative of the field's development, featuring comparable levels of loop complexity and lines of code (LoC). To further stress-test our approach, we introduced an original benchmark \textbf{pIp} that feature significantly more complex data structures and intricate control-flow patterns than existing sets.

Regarding the exclusion of SpecGen\cite{SpecGen} from our direct comparison, we note two primary reasons: first, a fundamental architectural mismatch exists as SpecGen targets different programming  and specification languages (Java/JML in SpecGen \emph{vs}. C/ACSL in our work), creating translation barriers that make direct, automated comparison infeasible. Second, SpecGen was evaluated on a custom dataset rather than the established AutoSpec benchmark used in our study and other baseline works. By grounding our evaluation in widely recognized datasets while adding more challenging cases, we ensure that our results are both comparable to previous research and indicative of performance on complex softwares.

\section{Related Work}
\label{sec:relatedwork}

\noindent\paragraph{\textbf{Traditional approach.}}
Traditional techniques for program specification and invariant inference can be broadly 
categorized into dynamic and static methods. Dynamic approaches infer specifications 
and invariants by analyzing runtime data collected during test execution. 
Daikon~\cite{daikon} executes test cases to gather program traces and produces a set 
of candidate invariants based on a predefined library of templates, which are subsequently 
filtered using the observed data. Sharma et al.~\cite{A_Data_Driven_Approach_for_Algebraic_Loop} 
employ a similar approach to generate candidate invariants, but additionally verify their 
soundness using SMT solvers. In contrast, DIG~\cite{dig} bypasses template-based filtering 
and instead directly infers invariants from execution traces by integrating techniques 
from equation solving, polyhedra construction, and theorem proving. Despite their 
practicality, dynamic methods are highly dependent on the coverage and quality of the 
test suites.

Static methods eliminate the dependency on test cases and explore a variety of techniques, 
including constraint solving~\cite{DBLP:conf/cav/ColonSS03,DBLP:journals/pacmpl/LiuFYSL22}, 
abstract interpretation~\cite{Abstract_Interpretation_A_Unified_Lattice_Model_for_Static_Analysis_of_Programs_by_Construction_or_Approximation_of_Fixpoints}~\cite{Automatic_Discovery_of_Linear_Restreints_among_Variables_of_a_Program}, 
abductive inference~\cite{Inductive_Invariant_Generation_via_Abductive_Inference}, 
Craig interpolation~\cite{FiB_Squeezing_loop,DBLP:conf/cade/GanDXZKC16,Nonlinear_Craig_Interpolant_Generation}, 
recurrence analysis~\cite{ISSACDeepakKapur,DBLP:journals/jsc/Rodriguez-CarbonellK07,DBLP:journals/pacmpl/CyphertK24}, 
shape analysis, and other methods~\cite{Backward_Symbolic_Execution_with_Loop_Folding}~\cite{SymInfer_Inferring_Numerical_Invariants_using_Symbolic_States}. 
Col{\'{o}}n \emph{et al}.~\cite{DBLP:conf/cav/ColonSS03} present the first complete 
invariant generation method based on Farkas' lemma and solves the non-linear invariant 
constraints via quantifier elimination, which is of double exponential time complexity. 
Some heuristics are proposed to promote the efficiency of solving the 
constraints~\cite{DBLP:journals/pacmpl/LiuFYSL22}. Cousot \emph{et al}.~\cite{Automatic_Discovery_of_Linear_Restreints_among_Variables_of_a_Program} 
utilize convex polyhedra along with transformations and widening operations to discover 
linear invariants based on abstract interpretation. However, due to the intrinsic nature 
of abstract interpretation, the efficiency of the analysis heavily depends on the choice 
of abstract domains, and moreover, the approach offers no theoretical guarantee on the 
accuracy of the generated invariants due to the precision loss in abstraction. 
Lin \emph{et al}.~\cite{FiB_Squeezing_loop} synthesize inductive invariants using Craig 
interpolants derived from reachability analyses, and nonlinear Craig interpolant generation 
was first studied in~\cite{Nonlinear_Craig_Interpolant_Generation}. The recurrence-based 
methods~\cite{ISSACDeepakKapur,DBLP:journals/jsc/Rodriguez-CarbonellK07,DBLP:journals/pacmpl/CyphertK24} 
derive recurrence relations from loops and solve these recurrences to obtain closed-form 
solutions as invariants. A key limitation of this approach, however, is restricted to 
the cases where the recurrences admit closed-form solutions or satisfy some strong conditions. 
Wu \emph{et al}.~\cite{Synthesizing_Invariants_for_Polynomial_Programs_by_Semidefinite_Programming} 
solve the strong and weak invariant synthesis problems of polynomial programs by combining 
invariant templates, semidefinite programming and constraint solving techniques. 
However, these traditional static methods are often not scalable and may require manual 
intervention. In contrast to these techniques, which are powerful yet often specialized 
for programs adhering to specific forms (e.g., certain types of loops or invariants), 
our approach leverages LLMs to achieve greater generality. In particular, some of these 
works focus primarily on theoretical challenges, and as a result, often exhibit limited 
efficiency and are difficult to apply directly to invariant generation at the program scale.

\noindent\paragraph{\textbf{Machine learning and SMT-based approach.}}
Machine learning techniques and Large Language Models have been increasingly used for 
synthesizing loop invariants. Recent advancements~\cite{iRank, Lemur, Code2Inv_a_Deep_Learning_Framework_for_Program_Verification, Loop_Invariant_Inference_through_SMT_Solving_Enhanced_Reinforcement_Learning, LLM_Generated_Invariants_for_Bounded_Model_Checking_Without_Loop_Unrolling}, 
particularly in deep learning and reinforcement learning, have spurred new approaches 
for invariant and specification generation. CODE2INV~\cite{Code2Inv_a_Deep_Learning_Framework_for_Program_Verification}, 
based on Counterexample-Guided Inductive Synthesis (CEGIS)~\cite{Counterexample_Guided_Inductive_Synthesis_Modulo_Theories}, 
utilizes graph neural networks to learn neural representations of program structures. 
While CODE2INV employs a reinforcement learning approach, it often faces challenges with 
limited performance and sparse rewards. To address these issues, LIPuS~\cite{Loop_Invariant_Inference_through_SMT_Solving_Enhanced_Reinforcement_Learning} 
establishes a clearer division of labor between the capabilities of RL and SMT solvers 
and introduces a novel two-dimensional reward mechanism to alleviate the sparse reward 
problem of CODE2INV. Clause2Inv~\cite{DBLP:journals/pacmse/CaoWXYWCM25} refines the 
guess-and-check paradigm into a generate-combine-check framework, employing an LLM to 
generate candidate clauses and a counterexample-driven combiner to synthesize them into 
loop invariants. However, these approaches are mostly confined to numeric programs and 
do not constitute an end-to-end invariant generation framework; instead, they produce 
invariants in SMT format. Moreover, they are goal-directed, i.e., they rely on pre-given 
verification goals to generate accurate specifications. Empirical comparisons on numeric 
benchmarks show that our tool achieves competitive performance, and in some cases even 
outperforms these methods.

\noindent\paragraph{\textbf{LLM-based approach}}
Pei \emph{et al}. \cite{Can_Large_Language_Models_Reason_About_Program_Invariants} demonstrates that fine-tuning pre-trained LLMs on code can effectively predict program invariants, with better performance than Daikon  if the set of available tests is limited. However, same as Daikon, it has no formal  guarantee for the validity of the generated invariants through formal verification. SpecGen~\cite{SpecGen} is a Java-oriented tool that employs a two-phase process to generate specifications: it first leverages LLM to produce invariant candidates, and then applies mutation operators with a weighted heuristic strategy to generate verifiable specifications when initial generation fails. Preguss \cite{Wang2025ATO} adopts a divide-and-conquer strategy that leverages potential runtime error assertions from static analysis to guide LLM-based synthesis of interprocedural specifications. AutoSpec~\cite{autospec} decomposes programs into smaller parts and uses carefully designed prompts including few-shot examples to guide an LLM in generating specifications, which are then verified using Frama-C. ACInv~\cite{Enhancing_Automated_Loop_Invariant_Generation_for_Complex_Programs_with_Large_Language_Models} utilizes static analysis to extract essential information about loops and data structures, which is then embedded into prompts for an LLM to generate candidate invariants and abstract predicates. We provide a detailed comparison in Section~\ref{sec:experiments} with ACInv~\cite{Enhancing_Automated_Loop_Invariant_Generation_for_Complex_Programs_with_Large_Language_Models} and AutoSpec~\cite{autospec}, both of which target C programs.


\section{Conclusion}
\label{sec:conclusion}

In this paper, we introduce a novel end-to-end framework for the automatic generation of C program specifications, including loop invariants and pre-/postconditions, by integrating symbolic execution, large language models (LLMs), and formal verification. Experimental results across diverse benchmarks demonstrate that our approach significantly outperforms existing methods in both invariant and functional specification generation. Furthermore, with the help of flattened memory model and inductive predicates, our tool  supports programs with complex data structures including structs, pointers and recursively defined types (e.g., single linked lists), as well as inductive operations. Notably, even without explicit verification goals, our method maintains competitive accuracy and consistently surpasses prior tools.

For future work, we aim to extend the framework to more complex recursive data structures and recursive functions, and to strengthen invariant generation for challenging cases such as nested loops, where correlations among multiple induction variables arise. In particular, we plan to enhance the current LLM-based approach by integrating it with symbolic sampling and targeted training strategies to improve precision of the invariant generation of nested loops.



\section{Data Availability}
All datasets, implementation code, experimental results, and supplementary materials including the appendix related to this work are publicly available at \url{https://github.com/anon-hiktyq/TOSEM2026-SESpec}.

\bibliographystyle{ACM-Reference-Format}
\bibliography{main}




\end{document}